\documentstyle[12pt]{article}
\date{}
\begin{document}

\title{Low lying $qqqq\bar q$ states in the baryon spectrum }

\author{C. Helminen$^1$
and D. O. Riska$^{1,2}$}
\maketitle

{\it $^1$ Department of Physics, 00014 University of Helsinki,
Finland}\par
{\it $^2$ Helsinki Institute of Physics, 00014 University of Helsinki,
Finland}\par

\vspace{0.8cm}

\centerline{\bf Abstract}
\vspace{0.5cm}

The coupling to light mesons leads to large widths and shifts of the energy
of the excited states in the baryon spectrum from the predictions of the
constituent quark model with three valence quarks. This coupling may be
modelled by admixtures of sea-quark $qqq(q\bar q)^n$ configurations
in the resonances. A schematic flavor and spin
dependent interaction model, similar to the one
which reproduces the low lying part of the
experimental baryon spectrum in the valence quark model,
is shown to bring the lowest $qqqq\bar q$ states with $L=1$ 
and positive parity below the
states with $L=0$. Because of the partial overlap with the
corresponding $qqq$ states this suggests that most of the low lying
states in the baryon resonance spectrum have substantial $qqqq\bar q$
admixtures. \\

\newpage

{\bf 1. Introduction}
\vspace{0.5cm}

The large color limit of QCD provides a simple and phenomenologically
satisfactory approach to analyze the
baryon spectrum \cite{Jenk}. This limit may either be realized through
the constituent quark model or through
topological soliton models, the generic version of which is the
Skyrme model \cite{Skyr}. Both approaches lead to a baryon spectrum, which
has the same ordering of positive and negative parity states as the
empirical spectrum, provided that the hyperfine interaction between the
constituent quarks in the quark model is assumed to have the flavor and
spin dependence characteristic of pion and vector meson exchange interaction
between quarks \cite{Kar, GlRi, Glo2}.
In particular, both approaches predict that the lowest excited states of
the nucleon and the $\Delta(1232)$ resonance are positive parity states,
which correspond to the N(1440),$\ {1\over2}^+$ and
$\Delta(1600),\ {3\over2}^+$ resonances, respectively, whereas the lowest
excited states in the spectrum of the $\Lambda$ hyperon have negative parity.\\

The description of these states is nevertheless quite different in the
two approaches. The constituent quark model with 3 valence quarks
describes them as a single quark
excitation to the excited S-state, whereas the Skyrme type model
describes them as collective breathing modes \cite{Bie}. The large
widths of these states do indeed indicate strong coupling to the open
mesonic channels, which in the quark model would be viewed as sea-quark
admixtures. In the meson sector similar sea-quark admixture was
noted by Jaffe \cite{Jaf0}, who noted that $(q\bar q)^2$ states
may have energies that are close to those of excited
$q\bar q$ states. In the baryon sector there is recent empirical evidence
for a dual character of at least the N(1440) \cite{Mor}.\\

The lowest negative parity states in the baryon spectrum are the
$\Lambda(1405)$, ${1\over2}^-$ and $\Lambda(1520),\ {3\over2}^-$ resonances,
which in the quark model with 3 valence quarks are described as a
flavor singlet multiplet \cite{Pai}.
The $qqq$ valence quark model with the pion exchange type hyperfine
interaction correctly predicts the low energy of this multiplet, but
cannot explain its spin-orbit splitting \cite{GlRi}. The bound state
version of the Skyrme model predicts both the low energy and the
spin-orbit splittings of this multiplet correctly, however \cite{Rho, Blom}.
This suggests that this multiplet should have a relatively large $qqqq\bar q$
component. The $\Lambda(1405)$ may even have a dominant $\bar K$N
component \cite{Dal}, and a separate $qqq$ component with a
$qqqq\bar q$ component that is strongly coupled to the $\bar K$N and
$\pi\Sigma$ systems \cite{Kim}.\\

The lowest negative parity state in the nucleon spectrum is the
N(1535),$\ {1\over2}^-$ state, the energy of which is correctly predicted
by the $qqq$ model. Yet the N(1535)$\to$ N$\gamma$ transition form factor
falls off at a slower rate with momentum transfer than the nucleon
form factor, a feature that would be very hard to describe by the
quark model with 3 valence quarks \cite{Bur}, in which it has to be
an $L=1$ state which a priori should be more spatially extended than the
$L=0$ nucleon. This feature, along with its peculiarly strong
$\eta$ decay branch, suggests that this state should have significant
sea-quark admixtures. The same observation concerning strong $\eta$
decay modes very near the threshold for $\eta$ decay also applies to the
$\Lambda(1670),\ {1\over2}^-$ and the $\Sigma(1750),\ {1\over2}^-$ states
\cite{PDG}.\\

Significant $qqqq\bar q$ admixtures in the baryon resonances should be expected
to occur where the energies and level splittings in the $qqqq\bar q$
states are close to those of the $qqq$ states. We here compare the
energies of the low
lying baryon resonances that are obtained with $qqq$ and $qqqq\bar q$
states using the same schematic flavor spin interaction, which is known
to yield a spectrum for the $qqq$ states which agrees well with the
empirical baryon spectrum up to $\sim$ 1700 MeV.\\

In this schematic model the confining interaction is described as
harmonic, mainly for the purpose of analytic integrability of the
Hamiltonian. The strength of the confining interaction for the $q \bar
q$ interaction and the $qq$ interaction is
suggested by the conventional color dependence
$\mbox{\boldmath$\lambda$}_i^C\cdot
\mbox{\boldmath$\lambda$}_j^C$ of the confining interaction. For the
hyperfine interaction between quarks we adopt the simple form

$$H_{\chi}=-C_{\chi}\sum_{i<j}^N\mbox{\boldmath$\lambda$}_i^F\cdot
\mbox{\boldmath$\lambda$}_j^F
\mbox{\boldmath$\sigma$}_i\cdot\mbox{\boldmath$\sigma$}_j\ ,\eqno(1.1)$$

\noindent where $C_{\chi}$ is a constant. The hyperfine interaction between
quarks and antiquarks is taken to be negligible, as suggested by the small
hyperfine splittings in
the empirical meson spectrum, with exception of the light pseudoscalar
octet mesons. In view of their small mass
the latter are better described as the Goldstone bosons of
the spontaneously broken approximate chiral symmetry of QCD than as
$q\bar q$ states.\\

If the constant $C_{\chi}$ in Eq. (1.1) is chosen so that both the
N(1440), ${1\over2}^+$ and N(1535), ${1\over2}^-$ resonances can be
described as $qqq$ states with additional $qqqq\bar q$ components
the lowest $qqqq\bar q$ states are states with
positive parity. The energy difference between the lowest lying negative
and positive parity $qqqq\bar q$ states is, due to different group
theoretical structures of the states, larger for the
nucleon-like states than for the $\Delta$-like (and  $\Omega^-$-like) states,
while the difference for the $\Lambda$-, $\Sigma$- and $\Xi$-like states is
much smaller. The low lying $qqqq\bar q$ states provide a discrete
approximation to the open $qqq$-meson channels that are the cause of the
large width of the  empirical low lying baryon resonances.
The most interesting feature of this rich spectrum are the
discrete low lying states, which in many cases overlap with the
corresponding low lying excited $qqq$ states, and thus suggest significant
$qqqq\bar q$ admixtures in several of the latter states. The higher
lying $qqqq\bar q$ states, on the other hand are so dense as
to almost form a near continuum of states.\\

The main result of this investigation for the structure of the nucleon
is that the $qqqq\bar q$ system has a well separated low lying
${1\over2}^+$ state, which overlaps with the N(1440), and supports the
view of substantial collectivity of that resonance. The $qqqq\bar q$
system also has a low lying  ${1\over2}^-$ - ${3\over2}^-$ doublet,
which similarly indicates that the corresponding
empirical states N(1535) and N(1520) have strong $qqqq\bar q$ admixtures.
\\

The lowest positive parity $qqqq\bar q$ state in the spectrum of the
$\Delta$ resonance is a triplet of ${1\over2}^+$, ${3\over2}^+$ and
${5\over2}^+$ states. The ${3\over2}^+$ state partly overlaps with the
empirical $\Delta(1600)$ resonance, and the low lying negative parity
states overlap with the empirical $\Delta(1620)$ and $\Delta(1700)$
states, thus suggesting substantial
sea-quark admixture in those states and a qualitative explanation
for the
large widths of these states.\\

In the case of the spectrum of the $\Lambda$-hyperon the outstanding
feature of the $qqqq\bar q$ spectrum is that in the present model the
lowest lying state is a ${1\over2}^+$ - ${3\over2}^+$ doublet state
but that it also has a well
separated low lying ${1\over2}^-$ state, which may suggest that the
$\Lambda(1405)$ is partly a 5-quark state (an $\bar K$N molecule)
\cite{Dal}.
About $\sim$ 65 MeV above this state lies a ${1\over2}^-$ - ${3\over2}^-$
multiplet which for the ${3\over2}^-$ overlaps with the
$\Lambda(1520),\ {3\over2}^-$ resonance.
The lowest lying positive and negative parity $qqqq\bar q$ states are
close in energy, and this difference is very sensitive to the value of
the parameter $C_{\chi}$. A slightly smaller $C_{\chi}$ would switch the
ordering of the lowest positive and negative $\Lambda$-like states, and also
of the $\Sigma$- and $\Xi$-like states,
while for the N-, $\Delta$- and $\Omega^-$-like states a positive parity state
would still be the lowest in energy.
In the case of the strange hyperons the predicted
$qqqq\bar q$ spectrum as expected corresponds well to the spectrum
predicted with the bound state version of the topological soliton model
\cite{Blom, Hor}.\\

This paper falls into 6 sections. In section 2 the Hamiltonian for
the $qqqq\bar q$ states, which is formed of a harmonic confining term
and a flavor-spin dependent hyperfine interaction, is presented along
with the symmetry classification of its eigenstates with $L=0$ and 1. In
section 3 the energies of the $qqqq\bar q$ states that have the flavor
quantum numbers of the nucleon and the $\Delta(1232)$ resonances are
calculated. The corresponding energies of the $qqqq\bar q$ states in the
spectrum of the $\Lambda$ and $\Sigma$ hyperons are given in section 4
and those of the $\Xi$ and $\Omega^-$ are given in section 5. Section 6
contains a concluding discussion.\\

\vspace{1cm}

{\bf 2. The $qqqq\bar q$ system}
\vspace{0.5cm}

A translationally invariant Hamiltonian model for the $qqqq\bar q$
system, the simplest version of which would be the harmonic oscillator
Hamiltonian, may be written as

$$H=\sum_{i=1}^5 {\vec p_i^{\ 2}\over2m_i} - {\vec P^{2}\over2M}
+ \sum_{i<j}^5 V_{conf}(r_{ij})+\sum_{i=1}^5 m_i\ .\eqno(2.1)$$

\noindent \noindent Here $m_i$ denotes the constituent masses of the quarks
(and the antiquark), and $\vec P$ and  $M$ are the total momentum and
mass of the $qqqq\bar q$ system respectively.
The confining potential $V_{conf}(r_{ij})$ will be taken
to have the form

$$V_{conf}(r_{ij})=-{3\over8}\mbox{\boldmath$\lambda$}_i^C\cdot
\mbox{\boldmath$\lambda$}_j^C(C\lbrack r_i-r_j\rbrack^2
+V_0)\ .\eqno(2.2)$$

\noindent In Eq. (2.2) the color dependence of the potential is
indicated by $\mbox{\boldmath$\lambda$}_i^C\cdot
\mbox{\boldmath$\lambda$}_j^C$, and $C$ and $V_0$ are constants.
The presence of negative constants such as $V_0$ is
required for a realistic description of the baryon
spectrum \cite{Glo2}. The above form for the confining potential is
chosen in order to achieve agreement with Regge phenomenology that
indicates that the string tension between quarks and antiquarks in
mesons is twice that between two quarks in a baryon.
Thus, with a confining interaction described by Eq. (2.2),
$<\mbox{\boldmath$\lambda$}_i^C\cdot\mbox{\boldmath$\lambda$}_j^C>=-8/3$
for a $qq$-pair in a baryon, while $<\mbox{\boldmath$\lambda$}_i^C\cdot
\mbox{\boldmath$\lambda$}_j^C>=-16/3$ for $q\bar q$ in mesons.
The situation for $qqqq\bar q$ states is, however, different, as has been
shown in Ref. \cite{Gen}. Due to the color symmetry structure of the $qqqq\bar
q$-system, $<\mbox{\boldmath$\lambda$}_i^C\cdot
\mbox{\boldmath$\lambda$}_j^C>=-4/3$ for both $qq$- and $q\bar q$-pairs.
It is thus possible to rewrite the Hamiltonian (2.1) as

$$H=\sum_{i=1}^5 {\vec p_i^{\ 2}\over2m_i} - {\vec P^{2}\over2M}
+ {1\over2}\sum_{i<j}^5(C\lbrack r_i-r_j\rbrack^2+V_0)+\sum_{i=1}^5 m_i\ .
\eqno(2.3)$$
\\

Note that the color dependence of the confining interaction
is conventional and mainly motivated by fact that the shell
spacing in the experimental meson spectrum is larger
($\sim$ 700 MeV) than it is in the experimental baryon
spectrum ($\sim$ 400 MeV). The recent quenched lattice
calculation of the effective confining interaction in
heavy $q\bar q$ and heavy $qqq$ systems suggests that
there is no color dependence in the confining interaction
\cite{Taka}.
The result above concerning the strength of the effective confining
interaction between an antiquark and a 4-quark system
is by the result of Ref. \cite{Gen} independent of the
color (in)dependence of the confining interaction.\\

If the constituent masses of the quarks (and the antiquark) are taken to
be equal the translationally invariant Hamiltonian
(2.3) may be rewritten as a sum of 4 uncoupled harmonic oscillator
Hamiltonians by the following change of variables:

$$\vec R={1\over\sqrt{5}}(\vec r_1+\vec r_2+\vec r_3 +\vec r_4 +
\vec r_5)\ ,\ \quad $$
$$\vec \xi_1={1\over\sqrt{2}}(\vec r_1-\vec r_2)\ ,\qquad\qquad
\qquad\qquad$$
$$\vec \xi_2={1\over\sqrt{6}}(\vec r_1+\vec r_2-2\vec r_3)\ ,\quad
\qquad\qquad\ $$
$$\vec \xi_3={1\over\sqrt{12}}(\vec r_1+\vec r_2+\vec r_3-3\vec r_4)\ ,
\qquad\quad$$
$$\vec \xi_4={1\over\sqrt{20}}(\vec r_1+\vec r_2+\vec r_3 +\vec r_4-
4\vec r_5)\ .\quad \eqno(2.4)$$

\noindent Here $\vec R$ is the c.m. coordinate. The resulting
uncoupled Hamiltonian then takes the form

$$\tilde H={1\over2m}\sum_{i=1}^4\eta_i^2+
{5\over2}C\sum_{i=1}^4\xi_i^2+
5V_0+5m\ .\eqno(2.5)$$

\noindent Here $\vec\eta_i$ is the momentum operator, which is
canonically conjugate to the operator $\vec\xi_i$
($\vec\eta_i=-i\vec\nabla_{\vec\xi_i}$). The Hamiltonian (2.5) describes
4 uncoupled harmonic oscillators, with the oscillator
frequency $\omega=\sqrt{5C\over m}$. Note that the
choice of variables (2.4) is suffices to reduce the
Hamiltonian (2.3) to a set of uncoupled oscillators
also in the more general case, where the string
tension is different for the different quark
pairs.\\

When one or more of the quarks in the $qqqq\bar q$ system
is strange, this Hamiltonian is only approximate. The
main effect of the difference between the constituent
masses of the light flavor and strange quarks is to shift
the energy upwards by multiples of the number of strange
quarks. This effect will be taken into account here.
The energy levels of the Hamiltonian (2.5) are

$$E_0=6\omega+N\omega+5V_0+5m\ ,\eqno(2.6)$$

\noindent where $N$ is the number of excited quanta of the oscillators.\\

The excited spectrum of the oscillator Hamiltonian is organized in
shells, the ground state shell of which ($N=0$) is formed of
states with negative parity because of the negative parity of the antiquark.
The lowest positive parity states occur in
the first excited P-shell and have excitation energies $\omega$, with a
quark or an antiquark in a p-state. Because of the negative parity of
the ground state band it is clear that the $qqqq\bar q$ states represent
excited and not ground state baryons. The key point is, however, that
the hyperfine interaction (1.1) is strong enough to bring the lowest
$L=1$ states down to or below the lowest states with $L=0$ in the spectra
of the nucleon and the $\Delta(1232)$ resonances, and therefore the
lowest $qqqq\bar q$ states form a band with positive parity states.
This situation is analogous to the description
of the spectrum of the strange hyperons in the bound state version of
the topological soliton model \cite{Blom, Hor}.\\

The matrix element of the hyperfine interaction $H_{\chi}$
can be calculated as \cite{Jaf,GlRi1}

$$<\lbrack f\rbrack^{SU(6)}\lbrack f\rbrack^{SU(3)}
\lbrack f\rbrack^{SU(2)}|\sum_{i<j}^N\mbox{\boldmath$\lambda$}_i\cdot
\mbox{\boldmath$\lambda$}_j
\mbox{\boldmath$\sigma$}_i\cdot\mbox{\boldmath$\sigma$}_j|
\lbrack f\rbrack^{SU(6)}\lbrack f\rbrack^{SU(3)}
\lbrack f\rbrack^{SU(2)}>$$
$$=4C_2^{(6)}-2C_2^{(3)}-{4\over3}C_2^{(2)}-8N\ ,
\eqno(2.7)$$

\noindent where $C_2^{(n)}$ is the matrix element of the quadratic Casimir
operator of $SU(n)$, with $n=$ 6, 3, and 2, and $N$ is the number of quarks,
i.e. 4 in the subsystem of four quarks. The matrix element $C_2^{(n)}$
is given by \cite{GlRi1}

$$C_2^{(n)}={1\over2}\lbrack f_1^{'}(f_1^{'}+n-1)+f_2^{'}(f_2^{'}+n-3)+\dots$$
$$\qquad\qquad +f_{n-1}^{'}(f_{n-1}^{'}-n+3)\rbrack
-{1\over2n}\left(\sum_{i=1}^{n-1}f_i^{'}\right)^2\ ,\eqno(2.8)$$

\noindent where
$f_i^{'}=(f_i-f_n)$, with $f_i$ being the length of the $i$th row
of the corresponding Young pattern.
The numerical values for $C_2^{(n)}$
are given in Table 1.\\

The state in which the 4 quarks are in their lowest s-states is
completely symmetric in the position coordinates ($\lbrack 4\rbrack_X$).
The Pauli principle demands that the corresponding color-flavor-spin
state be completely antisymmetric ($\lbrack 1111\rbrack_{CFS}$). This
symmetry is realized if the color state has the mixed symmetry $\lbrack
211\rbrack_C$ and therefore the flavor-spin state has the mixed symmetry
$\lbrack 31\rbrack_{FS}$. In Table 2 the matrix elements of the hyperfine
interaction Hamiltonian (1.1) are listed for all the combinations of
flavor and spin states, which have this mixed symmetry character.\\

The lowest state with $L=1$ has one quark in the p-state, and
consequently the symmetry of the orbital state is $\lbrack 31\rbrack_X$.
The corresponding color-flavor-spin state has accordingly to have the
mixed symmetry $\lbrack 211\rbrack_{CFS}$, and since the symmetry of the
color state is again $\lbrack 211\rbrack_C$, the possible symmetries are
$\lbrack 4\rbrack_{FS}$, $\lbrack 31\rbrack_{FS}$, $\lbrack 22\rbrack_{FS}$
and $\lbrack 211\rbrack_{FS}$.
The calculated matrix elements of the hyperfine interaction (1.1) for
states with these 4 flavor-spin symmetries are listed in Table 3. The
relative positions of these states with respect to those that have $L=0$
is obtained by addition of the orbital excitation energy
$\omega=\sqrt{5C\over m}$.
Since positive parity $L=1$ states can also be formed if the four quarks
are in their lowest s-states ($\lbrack 4\rbrack_X$) while the antiquark is
in a p-state the states with the four-quark flavor-spin symmetry
$\lbrack 31\rbrack_{FS}$  will be repeated at the energy
$\omega$ above the corresponding $L=0$ states.\\

In order to compare the calculated $qqqq\bar q$ spectrum with the
corresponding $qqq$ spectrum for the baryons, the value of the confining
constant $C$ is determined from the oscillator parameter in the three quark
baryon system, $\omega_0=\sqrt{6C\over m}$,
which in turn is determined by the empirical splitting between the
N(1440),$\ {1\over2}^+$ resonance and the nucleon to be $\omega_0=250$ MeV.
The splitting between states with the same flavor-spin symmetry in the
(lowest) $L=1$ and $L=0$ bands of the $qqqq\bar q$ system is then $\omega=
\sqrt{5\over6}\ \omega_0\approx 228$ MeV. The parameter $C_{\chi}$
in the hyperfine quark-quark interaction (1.1) is then adjusted to the
value $C_{\chi}= 21$ MeV to allow for $qqqq\bar q$ admixtures in the
lowest lying positive and negative parity nucleon resonance states.
With these parameter values the ground state of the $qqqq\bar q$ system will
be the $\lbrack 31\rbrack_X\lbrack 4\rbrack_{FS}\lbrack
22\rbrack_F\lbrack22\rbrack_S$ positive parity state \cite{Sta}, 
which then for the nucleon
mixes with the N(1440), ${1\over2}^+$ resonance.\\

Assessment of the possibility for significant admixture of $qqqq\bar q$
configurations into the $qqq$ states requires determination of the spin
flavor quantum numbers in the $qqqq\bar q$ multiplets for comparison
with those of the $qqq$ multiplet. The latter correspond well to the
measured quantum numbers of the states in the empirical baryon spectrum.\\

The isospin and strangeness content of the $qqqq\bar q$ states are most
simply analyzed by considering the $qqqq$ subsystem first, and subsequently
adding the antiquark. The flavor multiplet $\lbrack 4\rbrack_F$, which
corresponds to the $SU(3)$ flavor representation ${\bf 15}^{'}$, is then
combined with an antiquark in the flavor representation $\overline{\bf 3}$,
giving ${\bf 15}^{'}\times\overline{\bf 3}={\bf 10}+{\bf 35}$.
The two resulting representations are shown in $(I_3,Y)$ diagrams in Fig. 1,
where $I_3$ is the third component of the isospin $I$, and $Y$ is the
hypercharge, defined as the sum of the baryon number $B$ and the strangeness
number $S$. The $qqqq\bar q$ states in the decuplet that results from
$\lbrack 4\rbrack_F$ have the correct quantum numbers to describe mixing
with a three-quark decuplet.\\

The flavor multiplet $\lbrack 31\rbrack_F$, on the other hand,
can be described by the representation ${\bf 15}$ which, when combined
with an antiquark, gives ${\bf 15}\times\overline{\bf 3}={\bf 8}+{\bf 10}+
{\bf 27}$. These representations are shown in Fig. 2. The states in the
octet and decuplet of the resulting $qqqq\bar q$
states can in this case be
mixed with the three-quark flavor octet and decuplet states.
The $\lbrack 22\rbrack_F$ flavor multiplet  corresponds to the representation
$\overline{\bf 6}$, which yields $\overline{\bf 6}\times
\overline{\bf 3}={\bf 8}+\overline{\bf {10}}$ when
combined with an antiquark. These representations are shown in Fig. 3.
Mixing with a three-quark flavor octet state is thus possible.
Finally, the $\lbrack 211\rbrack_F$ flavor multiplet, described by the
representation ${\bf 3}$ will, when combined with an antiquark, yield
${\bf 3}\times\overline{\bf 3}={\bf 1}+{\bf 8}$ (Fig. 4) and
mixing with the three-quark flavor singlet and octet states,
respectively, is thus possible.\\

To summarize, the flavor decuplet states ($\Delta$, $\Sigma^{*}$, $\Xi^{*}$,
$\Omega^-$) can be derived from the $\lbrack 4\rbrack_F$ and $\lbrack
31\rbrack_F$  four-quark subsystems,  the flavor octet states (N,
$\Lambda$, $\Sigma$, $\Xi$) from $\lbrack 31\rbrack_F$,
$\lbrack 22\rbrack_F$ and $\lbrack 211\rbrack_F$, and, finally, the
flavor singlet state ($\Lambda^{'}$) from $\lbrack 211\rbrack_F$.
To derive some of the states, e.g. the nucleon from the four quark
symmetry $\lbrack 211\rbrack_F$, it is, however, necessary to assume
that the states contain $s\bar s$ states.
Since in this model the hyperfine splittings of the states are
determined by the flavor-spin structure
of the four-quark subsystem different energy states will therefore be denoted
below by their four-quark $\lbrack f\rbrack_{FS}\lbrack f\rbrack_F
\lbrack f\rbrack_S$ structure. \\

The total angular momentum $J$ of the $qqqq\bar q$ system can be
calculated for states of different spin symmetry in the four-quark
subsystem by first adding the spins of the four quarks to reach a state
$\lbrack f\rbrack_{S}$, and then adding the spin of
the antiquark (${\bf s={1\over2}}$) and, finally,
adding the orbital angular momentum if $L\ne0$.
The results for the different states are given in Table 4, where the
parity $P$ of the states has also been given. Note that in the $qqqq\bar q$
system the ground state configuration will have $P=-1$ due to the
presence of a $q\bar q$ pair and, subsequently, the first excited state
will have positive parity.\\

\vspace{1cm}

{\bf 3. 5 quark components in the nucleon and $\Delta$ resonance states}
\vspace{0.5cm}

States with the quantum numbers of the nucleon resonances ($\lbrack
21\rbrack_F$) appear in the  $qqqq\bar q$ states that belong to the
flavor multiplets $\lbrack 31\rbrack_F$, $\lbrack
22\rbrack_F$, and $\lbrack 211\rbrack_F$ of the four-quark
subsystem. The symmetry $\lbrack 211\rbrack_F$ has, however, to
contain a strange quark, and to have total strangeness zero the 
antiquark has to be
$\bar s$ in this $qqqq\bar q$ state. Nucleon-like states derived form the
states with four-quark symmetries $\lbrack 31\rbrack_F$ and
$\lbrack 22\rbrack_F$ may be both states without and with hidden strangeness.
Among the $qqqq\bar q$ states with zero strangeness that can mix with
the nucleon there are thus many states with $s\bar s$
pairs, and which therefore represent strangeness components of the
nucleon. Experimental signatures for such
$s\bar s$ components in the nucleon have been a topic of intense
experimental interest recently \cite{SAM, HAP}. Here it has been assumed
that the explicit $qqqq\bar q$ configurations have energies that lie above
the lowest excited nucleon resonance energy. Consequently signatures
of these configurations are expected to be most visible in the decay
patterns of the nucleon. The main effect of $SU(3)_F$ breaking is a shift
of the energies of these $s\bar s$ configurations upwards by $\sim 2\delta m$,
where $\delta m$ is the difference between the constituent masses of the
strange and light flavor quarks. We shall take $\delta m$ to be
$\delta m=m_s-m_u=120$ MeV \cite{GlRi}. The $L=0$ state
$\lbrack 31\rbrack_{FS}\lbrack 211\rbrack_F\lbrack 22\rbrack_S$ that
otherwise would be the lowest lying negative parity state, will for the
nucleon be lifted up $2\delta m$ due to the flavor symmetry structure
$\lbrack 211\rbrack_F$ of the four-quark system.\\

We interpret the $qqqq\bar q$ states as sea-quark admixtures of the
excited $qqq$ states, and therefore, as mentioned earlier, take the energy
of the lowest nucleon resonance, the N(1440), as the reference point for the
energy of the positive parity ground state of the
$qqqq\bar q$ system. The N(1440), ${1\over 2}^+$ state has a complex
structure and is expected to have large sea-quark and possible more
exotic components \cite{Bur1}. The expression for the mass of the lowest
positive parity $qqqq\bar q$ state in the nucleon spectrum is

$$E\lbrace\lbrack 4\rbrack_{FS}\lbrack 22\rbrack_{F}\lbrack
22\rbrack_{S}\rbrace=7\omega-28C_{\chi}+5V_0+5m\ .
\eqno(3.1)$$

\noindent Using the parameter values above and taking $m=340$ MeV and
$V_0=-269$ MeV the energy of this state is 1365
MeV, which is the empirical value for the real part of the
pole position of the N(1440) resonance. 
The real part of the pole position is the appropriate
energy to which compare quark models, as it takes into
account the shift caused by the coupling to the open
mesonic channels \cite{Hohl}.
Smaller values for $V_0$ would
shift the energy of the lowest positive parity $qqqq\bar q$ state,
and consequently also the energy of other $qqqq\bar q$ states,
upwards.\\

The symmetry classification of those $qqqq\bar q$ states which have the
same quantum numbers as the nucleon resonances, which are predicted to
have energies below 1900 MeV, along with their estimated
energies are listed in Tables 5 and 6 for the states with negative and
positive parity respectively. Since there is no hyperfine interaction
between the quarks and antiquarks in the present model, states with the
same four-quark symmetry as the ground state but with an antiquark in a
p-state will repeat themselves in the positive parity spectrum
at the same energy as states with one quark in a p-state and with
similar flavor-spin symmetry in the $L=1$ band. For convenience the
corresponding $qqqq\bar q$ states, which describe $\Delta$ resonance
excitations are listed in the same tables. There are no nucleon resonances
with positive parity in the $\lbrack4\rbrack_F$ multiplet, and no $\Delta$
resonances in the flavor multiplets $\lbrack22\rbrack_F$ and
$\lbrack211\rbrack_F$.\\

In Fig. 5 the presently known empirical spectrum
of nucleon resonances with energy between 1.4 GeV and 2.5 GeV and known
spin-parity $J^P$ is shown for $J\le{7\over2}$.
In Tables 5 and 6 those known nucleon resonances, which may have significant
$qqqq\bar q$ components with the symmetry structure indicated are also
listed. Since the empirical SD shell of nucleon and $\Delta$ resonances
is not yet complete, we have indicated the probable $qqq$ states, which
have corresponding $qqqq\bar q$ admixtures, where the empirical
confirmation is still lacking. For the most part there is a good
correspondence between the predicted low lying $qqqq\bar q$ states and the
empirical nucleon spectrum in the negative parity sector. In the
positive parity sector all low lying empirical states lie close to
predicted $qqqq\bar q$ states, but there are, however, some low lying
$qqqq\bar q$ states that are not seen in the nucleon spectrum, especially
in the ${3\over2}^+$ sector, the lowest state of which empirically is the
N(1720) resonance.\\

States with quantum numbers of the $\Delta$ resonances ($\lbrack
3\rbrack_F$) are contained in the $\lbrack 4\rbrack_F$ and $\lbrack
31\rbrack_F$ flavor multiplets of the four-quark subsystem.
Some of these states may also contain hidden strangeness, i.e. $s\bar
s$-pairs. Such states then lie $2\delta m$ above corresponding states
without $s\bar s$-pairs.
The calculated spectrum of states with $J^P$ up to
${7\over2}^+$ and energy up to 2.5 GeV is shown in Fig. 6 along with
the presently empirically known $\Delta$ resonances. The numerical
values of the calculated energies of the negative and positive parity
$\Delta$ resonance states with calculated energy below 1900 MeV are
listed in Tables 5 and 6 respectively. In the column ''emp'' those empirically
known and predicted $qqq$ $\Delta$-resonance states, which are likely to
have corresponding $qqqq\bar q$ admixtures are also indicated.
As can be seen from Fig. 6 several of the empirical positive parity states
can be found close to corresponding $qqqq\bar q$ states, e.g.
the $\Delta(1600),\ {3\over2}^+$ resonance.
The empirical positive parity states $\Delta(1910),\ {1\over2}^+$,
$\Delta(1920),\ {3\over2}^+$ and $\Delta(1905),\ {5\over2}^+$
may have admixtures of the $\lbrack 31\rbrack_{FS}\lbrack 31\rbrack_{F}
\lbrack 31\rbrack_{S}$ multiplet, while the ''one star'' resonance
$\Delta(1750)$, ${1\over2}^+$ seems to have admixtures of several $qqqq\bar q$
states that are predicted to have energies around 1745 - 1785 MeV.
In the negative parity sector the $\Delta(1620),\ {1\over2}^-$ resonance
seems to have a substantial admixture of a
$\lbrack 31\rbrack_{FS}\lbrack 31\rbrack_{F}\lbrack 31\rbrack_{S}$
multiplet at 1613 MeV, and maybe also some admixture of a
$\lbrack 31\rbrack_{FS}\lbrack 31\rbrack_{F}\lbrack 22\rbrack_{S},
\ {1\over2}^-$ singlet state at $\sim$ 1560 MeV. The $\Delta(1700),
\ {3\over2}^-$ resonance (with the real part of the pole position at 1660 MeV)
also seems to have admixtures of several $qqqq\bar q$ states in an energy
range of about 1610 - 1780 MeV.\\

\vspace{1cm}

{\bf 4. 5 quark components in the $\Lambda$ and $\Sigma$ resonances}
\vspace{0.5cm}

The $qqqq\bar q$ states with strangeness $-1$ contain at least one
strange quark, and accordingly their spectrum begins at a level that is
shifted about $\delta m$ above the corresponding level for
non-strange $qqqq\bar q$ states. Combination of the $qqqq$ flavor multiplets
$\lbrack 31\rbrack_F$, $\lbrack 22\rbrack_F$ and $\lbrack
211\rbrack_F$ with an antiquark leads to states with the quantum numbers
of the $\Lambda$ hyperon resonances.  Of these only the
$\lbrack 211\rbrack_F$ multiplet can combine with an antiquark to form a
flavor singlet state. The states derived from the four-quark symmetries
$\lbrack 31\rbrack_F$ and $\lbrack 211\rbrack_F$ are states both with
one $s$ quark, and with $ss\bar s$-combinations, while from the
four-quark symmetry $\lbrack 22\rbrack_F$ only $\Lambda$-like states with one
strange quark can be derived. The states containing $ss\bar
s$-combinations lie $2\delta m$ above the corresponding states with one
$s$-quark.\\

In Tables 7 and 8 we list the calculated energies of the $\Lambda$
hyperon resonances in the $qqqq\bar q$ spectrum up to 2 GeV, along with
their symmetry character. The empirically known $\Lambda$ hyperon
resonances, which have similar energies and the same quantum numbers are
also indicated. These are the states, which are expected to have
correspondingly strong $qqqq\bar q$ admixtures. The calculated
$qqqq\bar q$\ \ $\Lambda$ hyperon spectrum up to 2.5 GeV and
$J^P={7\over2}^-$ with $L=0$ and 1 is shown in Fig. 7.\\

The outstanding feature of the $\Lambda$ hyperon spectrum that can be
formed of 5 quark states is the appearance of both a low lying ${1\over2}^-$
flavor singlet state and a low lying ${1\over2}^+$ - ${3\over2}^+$
doublet state. The former state may in this model give a small
$qqqq\bar q$ contribution the low lying
isolated $\Lambda(1405)$ state, which conventionally has been described
as a $\bar K$N molecule \cite{Dal} or in other words - a 5 quark
state.\\

Above this low lying flavor singlet ${1\over2}^-$ state are
found several ${1\over2}^-$ - ${3\over2}^-$ $qqqq\bar q$ doublets,
which are likely to be mixed into the $\Lambda(1670),\ {1\over2}^-$ and
$\Lambda(1690),\ {3\over2}^-$ resonances, and in the case of the lowest
negative parity doublet also probably with the $\Lambda(1520),\ {3\over2}^-$
state. Above these doublets further appear a singlet ${1\over2}^-$ state,
and several ${1\over2}^-$ - ${3\over2}^-$ doublets, which are likely to
be mixed into the empirical $\Lambda(1800),\ {1\over2}^-$,
and an expected but still ''missing'' ${3\over2}^-$ state near 1800 MeV.
There is also a ${3\over2}^-$ - ${5\over2}^-$ doublet, which is strongly
mixed into the $\Lambda(1830),\ {5\over2}^-$.
\\

The lowest positive parity $qqqq\bar q$ state in the spectrum of the
$\Lambda$ hyperon is the ${1\over2}^+$ - $\ {3\over2}^+$ multiplet
$\lbrack 4\rbrack_{FS}\lbrack 22\rbrack_{F}\lbrack 22\rbrack_{S}$,
which is found to have an energy of 1485 MeV. This multiplet is not seen
empirically. It may, however, form a small $qqqq\bar q$ admixture in the
empirical $\Lambda(1600),\ {1\over2}^+$ resonance.
About 140 MeV above this $qqqq\bar q$ state lies the 5 quark multiplet
$\lbrack 4\rbrack_{FS}\lbrack 31\rbrack_{F}\lbrack 31\rbrack_{S}$,
which has states with $J^P={1\over2}^+,\ {3\over2}^+$ and ${5\over2}^+$.
The first one of these is probably  mixed into the
$\Lambda(1600)$. Above this multiplet
appear several multiplets that may be mixed into the $\Lambda(1810),\
{1\over2}^+$, $\Lambda(1890),\ {3\over2}^+$ and $\Lambda(1820),\
{5\over2}^+$ resonances.\\

The spectrum of the $\Sigma$ hyperon does in all versions of the quark
model have the same structure as the spectra of the nucleon and the
$\Delta(1232)$ combined. Hence all the $qqqq$ flavor multiplets
$\lbrack 4\rbrack_F$, $\lbrack 31\rbrack_F$, $\lbrack 22\rbrack_F$ and
$\lbrack 211\rbrack_F$ can be combined with an antiquark to states that
have the quantum numbers of the $\Sigma$ resonances,
all but the states derived from the symmetry structure $\lbrack
211\rbrack_F$ containing one $s$-quark and $ss\bar s$-combinations.
States derived from the symmetry structure $\lbrack 211\rbrack_F$
contain only states with one $s$-quark and no $ss\bar s$-combinations.
In Tables 7 and 8
we list the energies and symmetry character of those $qqqq\bar q$
$\Sigma$ hyperon resonances that in the present model have energies
below 2 GeV. In Fig. 8 the $qqqq\bar q$\ \ $\Sigma$ states with
$J^P\le{7\over2}^+$ and $L=0$, 1 have been plotted for energies
up to 2.5 GeV.\\

The identification of which $\Sigma$ hyperon resonances should have
strong 5 quark admixtures is difficult because of the still very
incomplete empirical information of even the low energy part of the
$\Sigma$ hyperon spectrum. Whether e.g. the ''one-star'' resonance
$\Sigma(1480)$ is a real resonance, and the $qqqq\bar q$ analog of the
low lying $\Lambda(1405),\ {1\over2}^-$ state is not known.
On the other hand, in the present model there is a low lying positive
parity ${1\over2}^+$ - ${3\over2}^+$ multiplet with the energy 1485 MeV.
Similarly, the quantum numbers of the $\Sigma(1560)$ remain unknown.
If indeed the $\Lambda(1405)$ is partly a 5 quark state, it would be
natural to expect the $\Sigma(1560)$ to be the analog of the $\Lambda(1405)$,
and thus that it has $J^P={1\over2}^-$. This is indeed what the structure
of the $qqqq\bar q$ spectrum shown in Table 7 indicates. This would also
explain why the usual quark model description of the baryons as 3 quark
states cannot predict sufficiently low energies for the $\Lambda(1405)$
and $\Sigma(1560)$ \cite{GlRi}. The $\Sigma(1480)$(*) may then, on the
other hand, contain admixtures of the lowest lying positive parity
$qqqq\bar q$ state, thus having $J^P={1\over2}^+$ or $J^P={3\over2}^+$.
\\

Several of the negative parity $qqqq\bar q$\ \ $\Sigma$ hyperon resonance
states that lie above the lowest
$\lbrack 31\rbrack_{FS}\lbrack 211\rbrack_F\lbrack 22\rbrack_S$ state
have energies that are close to empirically established negative
parity states (Table 7). Thus the $\Sigma(1620)$, ${1\over2}^-$
''two-star'' state lies close to the predicted energy of both the
$\lbrack 31\rbrack_{FS}\lbrack 211\rbrack_F\lbrack 31\rbrack_S$ and the
$\lbrack 31\rbrack_{FS}\lbrack 22\rbrack_F\lbrack 31\rbrack_S$
$qqqq\bar q$ multiplets, and may be expected to have a substantial 5
quark component, while the former state is also probably mixed with the
$\Sigma(1580)$, ${3\over2}^-$ (''two-star'') state.
Similarly the $\Sigma(1670)$, ${3\over2}^-$ lies
close to the predicted energy of the
$\lbrack 31\rbrack_{FS}\lbrack 22\rbrack_F\lbrack 31\rbrack_S$
multiplet, and may also get a small $qqqq\bar q$ admixture from the
$\lbrack 31\rbrack_{FS}\lbrack 31\rbrack_F\lbrack 31\rbrack_S$
multiplet, and the $\Sigma(1690),\ {1\over2}^-$ (''two-star'') state
has an energy close to that of the
$\lbrack 31\rbrack_{FS}\lbrack 31\rbrack_F\lbrack 22\rbrack_S$ singlet state.
The predicted energy for the
$\lbrack 31\rbrack_{FS}\lbrack 31\rbrack_F\lbrack 31\rbrack_S$
multiplet lies close to the $\Sigma(1750),\ {1\over2}^-$ state, while
the $\Sigma(1940),\ {3\over2}^-$ is close in energy to several $qqqq\bar q$
multiplets. Consequently these resonances may very well be expected to have
substantial 5 quark components.\\

The lowest $qqqq\bar q$\ \ $\Sigma$ hyperon state with positive parity is
the $\lbrack 4\rbrack_{FS}\lbrack 22\rbrack_{F}\lbrack 22\rbrack_{S}$
multiplet (Table 8). In the present model its energy is found to be at
1485 MeV, which is 175 MeV below the $\Sigma(1660),\ {1\over2}^+$
resonance. Whether this state form an admixture in the ''one-star''
$\Sigma(1480)$ state is not clear, but it may in any case form a small
admixture in the $\Sigma(1660)$, in analogy with the corresponding admixture
in the $\Lambda(1600)$. The following state above this $qqqq\bar q$ multiplet
is the $\lbrack 4\rbrack_{FS}\lbrack 31\rbrack_{F}\lbrack 31\rbrack_{S}$
multiplet, which may form a significant admixture in the
$\Sigma(1660),\ {1\over2}^+$, as its energy is found to be
$\sim$ 1625 MeV. Above that lies a $qqss\bar s$ state  and a $qqqs\bar
q$ state with energies of 1725 MeV and 1737 MeV, respectively. Also
these states may form admixtures in the $\Sigma(1660),\ {1\over2}^+$,
and in the ''one-star'' $\Sigma(1770),\ {1\over2}^+$ resonance.
Above these are several $qqqq\bar q$ multiplets,
which in the present model form significant admixtures in the one- and
two-star states $\Sigma(1840),\ {3\over2}^+$ and
$\Sigma(1880),\ {1\over2}^+$, respectively, as well as states close in
energy to the empirical $\Sigma(1915),\ {5\over2}^+$ resonance.\\

\vspace{1cm}

{\bf 5. $qqqq\bar q$ components in the $\Xi$ and $\Omega^-$ resonances}
\vspace{0.5cm}

The $\Xi$ hyperons have strangeness $-2$ and therefore contain at least
two strange constituent quarks. States derived from the
$\lbrack 211\rbrack_{F}$ and $\lbrack 22\rbrack_{F}$ symmetries contain
only two $s$-quarks, while states derived from the $\lbrack4\rbrack_{F}$
and $\lbrack 31\rbrack_{F}$ symmetries have both states with two
$s$-quarks and $sss\bar s$-combinations.
Their spectrum, while otherwise similar to that of the $\Sigma$ hyperon
both in the 3 quark and 4 quark + 1 antiquark model, with the exception
mentioned above, therefore begins at an energy $2\delta m=2(m_s-m_u)$
above that of the nucleon and $\delta m$ above that of the $\Sigma$
hyperon spectrum.\\

The experimental spectrum of the $\Xi$ hyperon is rich, but poorly
understood, in fact of the resonances above the ground state band only
the $\Xi(1820)$ is known to be a ${3\over2}^-$ state. This makes it
impossible to make definite suggestions for which $\Xi$ resonances may
have significant $qqqq\bar q$ configurations.
The calculated $qqqq\bar q$\ \ $\Xi$ hyperon spectrum up to 2.5 GeV
and $J^P={7\over2}^-$ with $L=0$ and 1 is shown in Fig. 9 and the
numerical values for the energies for states up to 2.1 MeV are given in
Tables 9 and 10.\\

The lowest in energy of the $qqqq\bar q$\ \ $\Xi$ resonances with
negative parity is the
$\lbrack 31\rbrack_{FS}\lbrack 211\rbrack_{F}\lbrack 22\rbrack_{S}\
{1\over2}^-$ state, which in the present model is found to be at $\sim$ 1730
MeV, and above this is found a $qqqq\bar q$ ${1\over2}^-$ -
${3\over2}^-$ multiplet at 1685 MeV. If the $\Xi(1690)$ resonance
is confirmed as a ${1\over2}^-$ or a ${3\over2}^-$ state it is natural to
expect that it has a significant $qqqq\bar q$ component.
Also the negative parity $\Xi(1820)$ resonance may have admixtures of
several 5 quark states. As there is a negative parity $qqqq\bar q$ state
close in energy to the $\Xi(1950)$ this state may then be either a
${3\over2}^-$ or a ${5\over2}^-$ state with a significant $qqqq\bar q$
component.\\

The lowest lying positive parity
$qqqq\bar q$ state in the $\Xi$ spectrum is the ${1\over2}^+$ -
${3\over2}^+$ multiplet $\lbrack 4\rbrack_{FS}\lbrack 22\rbrack_{F}\lbrack
22\rbrack_{S}$. This is predicted to have an energy of 1605 MeV, which
is suggestively close to the one-star state $\Xi(1620)$, which is then likely a
positive parity state. If so the pattern of the other flavor sectors of
finding $qqqq\bar q$ states with positive parity repeats itself also in
the $\Xi$ spectrum. The $\Xi(1620)(*)$ should in any case be expected to
have a substantial $qqqq\bar q$ component. The $\Xi(2030)$, with
$J^P\le{5\over2}^?$ could also be a positive parity state, with either
$J^P={5\over2}^+$ or ${7\over2}^+$, containing significant $qqqq\bar q$
admixtures.\\

The symmetry structure of the $sssq\bar q$ states in the spectrum of the
$\Omega^-$ hyperon is the same as that of the $qqqq\bar q$ states in the
spectrum of the $\Delta(1232)$ resonance. From the $\lbrack 4\rbrack_{F}$
symmetry also states containing only $s$-quarks may be derived, while
the symmetry $\lbrack 31\rbrack_{F}$ implies only states that contain
three $s$-quarks. This spectrum should begin at
an energy 3$\delta m$ above that of the corresponding $\Delta$ spectrum.
The symmetry structure of the negative and positive parity $sssq\bar q$\
\ $\Omega^-$ resonances, with energies that are predicted to fall below
2.4 GeV is listed in Tables 11 and 12 respectively. In the absence of
empirical quantum number assignments for hitherto found $\Omega^-$
resonances it requires pure speculation to indicate which type of
$sssq\bar q$ admixture those states are likely to have.
In Fig. 10 the calculated $qqqq\bar q$\ \ $\Omega^-$ states with
$J^P\le{7\over2}^+$ and $L=0$, 1 have been plotted for energies
up to 2.5 GeV.\\

Both the negative and positive parity sectors of the excited $\Omega^-$\
\ $sssq\bar q$ spectrum begin, according to the present model, at around
1.9 GeV, with the positive parity multiplet $\lbrack 4\rbrack_{FS}
\lbrack 31\rbrack_{F}\lbrack 31\rbrack_{S}$ falling somewhat lower than
the corresponding negative parity multiplet $\lbrack 31\rbrack_{FS}
\lbrack 31\rbrack_{F}\lbrack 22\rbrack_{S}$. The corresponding 3
valence quark model for the $\Omega^-$ in contrast
suggests that the lowest excited
$\Omega^-$ state is the negative parity $\lbrack 21\rbrack_{FS}
\lbrack 3\rbrack_{F}\lbrack 21\rbrack_{S}$\ \ ${1\over2}^-$ - ${3\over2}^-$
multiplet, with an energy in the range 1950 - 1990 MeV 
\cite{GlRi,DanC}. The lowest
$sssq\bar q$ resonance with positive parity in the present model falls at
1865 MeV, while the lowest negative parity state lies at $\sim$ 1920 MeV.
The low lying part of the $sssq\bar q$\ \ $\Omega^-$ spectrum does in
any case resemble the corresponding excited spectrum of the 3 valence
quark model. Hence it is natural to expect considerable
sea-quark admixture in the spectrum of the $\Omega^-$ hyperon.\\

The only well established $\Omega^-$ resonance is the $\Omega(2250)^-$.
In the 3 valence quark model this has to be a positive parity state.
In the present model the positive parity state $\lbrack 31\rbrack_{FS}
\lbrack 31\rbrack_{F}\lbrack 31\rbrack_{S}$ has an energy of
$\sim 2200$ MeV. This would then suggest
that this resonance is likely to be partly a 5 quark state.\\

\vspace{1cm}

{\bf 6. Discussion}
\vspace{0.5cm}

Besides the 5 quark states that have been considered above, there will
also exist exotic $qqqq\bar q$ states with quantum numbers that
are excluded for
pure $qqq$ configurations. Such exotic states
can be formed from the flavor representations
${\bf 35}$, ${\bf 27}$ and $\overline{\bf 10}$, the four-quark flavor
substructure of which have the symmetries $\lbrack 4\rbrack_F$,
$\lbrack 31\rbrack_F$ and $\lbrack 22\rbrack_F$, e.g. states with
strangeness $S=1$ and isospin $I=2,1$ and $0$. The ${\bf 35}$
representation also includes other ''exotic'' states, such as
$(S,I)=\ (0,{5\over2}),\ (-1,2),\ (-2,{3\over2}),\ (-3,1)$ and
$(-4,{1\over2})$. The representation ${\bf 27}$, on the other hand,
contains ''exotic'' states with $(S,I)=\ (1,1),\ (-1,2),\ (-2,{3\over2})$
and $(-3,1)$ and, finally, the $\overline{\bf 10}$ representation
contains the ''exotic'' states $(S,I)=(1,0)$ and $(-2,{3\over2})$.
Baryon states with $S=+1$, with the structure $qqqq\bar s$, where $q$
represents $u$ or $d$ quarks, are the strange analogs of the $C=-1$
pentaquarks, which may be stable against strong decay \cite{Sta, Sco,
Gig, Ash}. Strange pentaquarks in contrast do decay strongly.\\

The conventional view that the baryon resonances should be described as
3 quark states is mainly due to the remarkably successful description
that the 3 valence quark model provides for not only the energies, but
the magnetic moments and axial couplings as well of the ground state
baryons in all flavor sectors of the baryon spectrum. No such simple
description has yet been found for the baryon resonances, although
the spectrum itself can be qualitatively described by
interaction models with the form considered here. As
noted above there are strong empirical and phenomenological indications
for large sea-quark admixtures in the low lying negative and positive
parity states. The main result of the present study is that the
chiral constituent quark model, a quark model with a flavor-spin dependent
hyperfine interaction, will bring the lowest $L=1$ positive parity
$qqqq\bar q$ states, with quantum numbers that correspond well with the
lowest lying positive parity baryon resonances above the ground state band,
below the lowest $L=0$ negative parity $qqqq\bar q$ states. Although the
present study is based on a somewhat schematic model for the flavor-spin
dependent hyperfine interaction, it nevertheless suggests that most of
the lowest lying baryon resonances in all flavor sectors of the
baryon spectrum have strong sea-quark components of the form 
$qqqq\bar q$. What is really needed to understand the spectrum of baryon
excitations in the framework of this paper is a model for the mixing 
between ordinary 3-body excitations and these $qqqq\bar{q}$ states.\\

\vspace{1cm}

{\bf Acknowledgment}
\vspace{0.5cm}

C.H. thanks the Arvid and Greta Olin Foundation for a stipend.
Research supported in part by the Academy of Finland through contract 43982.

\newpage

\centerline{\bf Figure Captions}
\vspace{0.5cm}

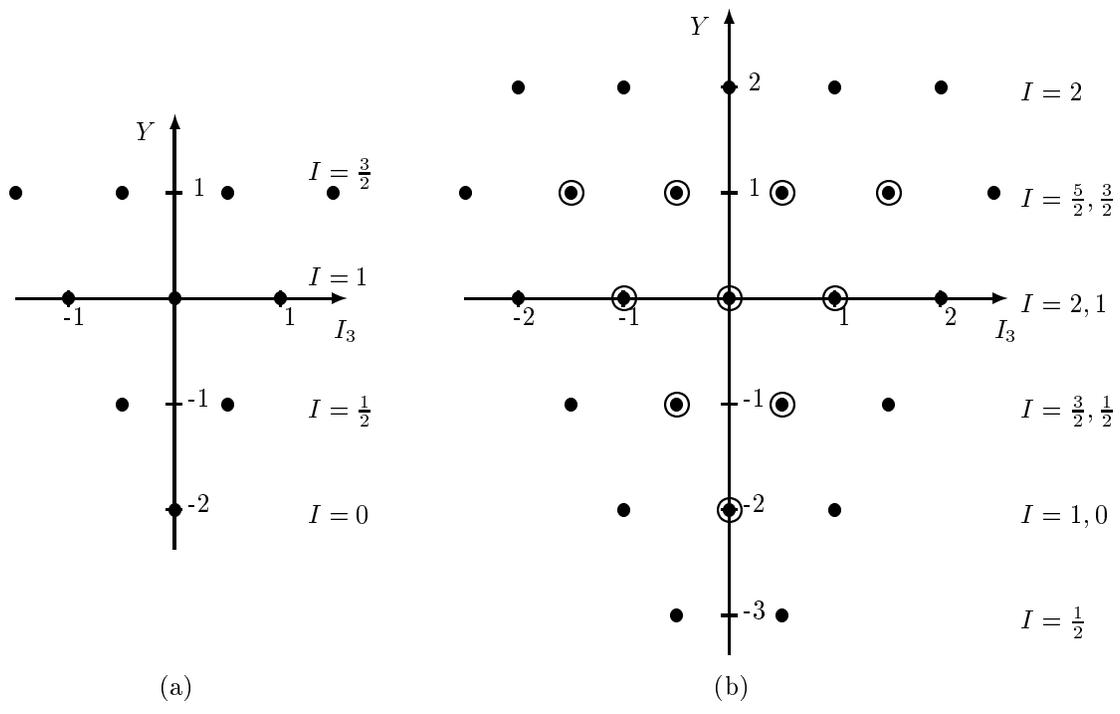
Figure 1: The representations (a) ${\bf 10}$ and (b) ${\bf 35}$
of ${\bf 15}^{'}\times\overline{\bf 3}$ in flavor $SU(3)$. Here $Y$ is the
hypercharge, $I$ isospin and $I_3$ the third component of the isospin.\\

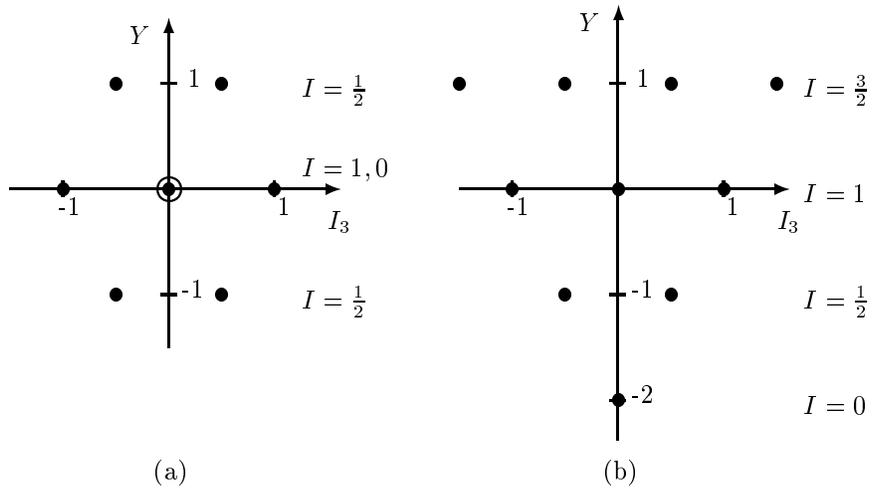
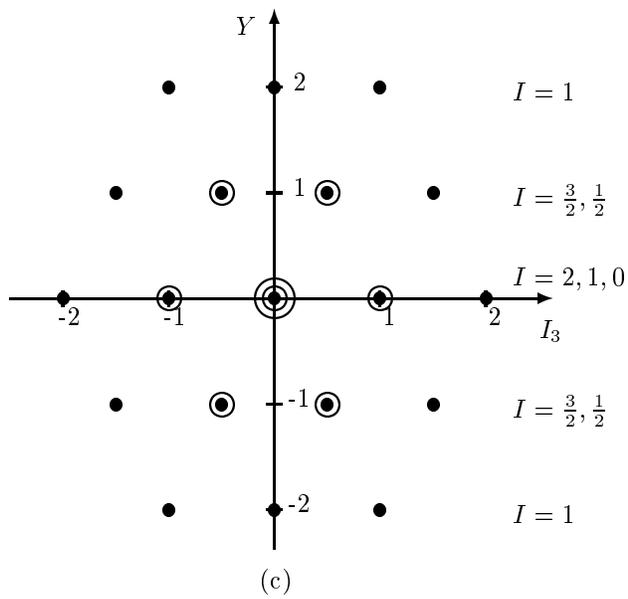
Figure 2: The representations (a) ${\bf 8}$, (b) ${\bf 10}$ and
(c) ${\bf 27}$ of ${\bf 15}\times\overline{\bf 3}$ in flavor $SU(3)$.
Here $Y$ is the hypercharge, $I$ isospin and $I_3$ the third component
of the isospin.\\

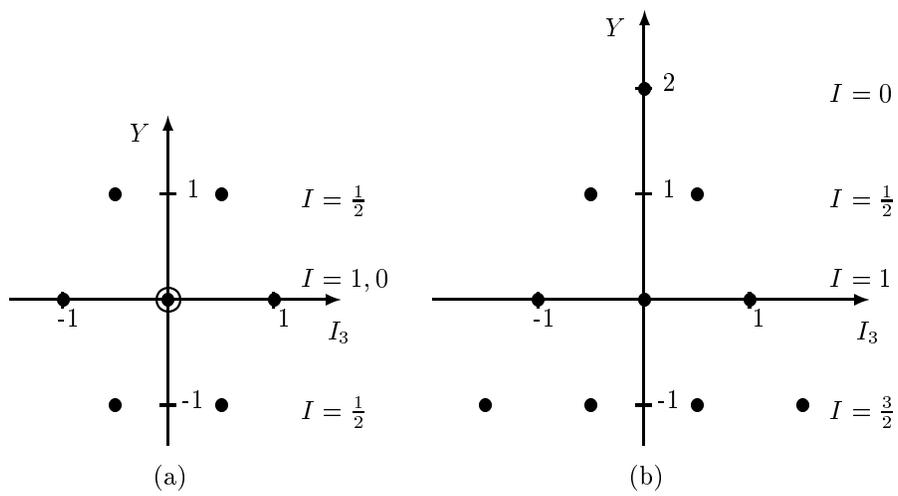
Figure 3: The representations (a) ${\bf 8}$ and (b) $\overline{\bf 10}$ of
$\overline{\bf 6}\times\overline{\bf 3}$ in flavor $SU(3)$. Here $Y$ is the
hypercharge, $I$ isospin and $I_3$ the third component of the isospin.\\

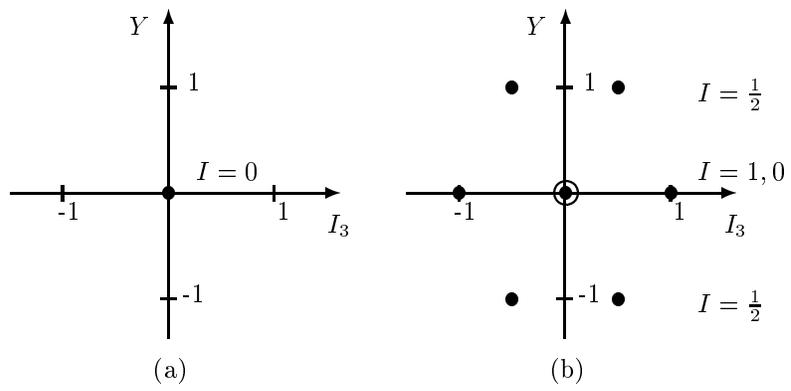
Figure 4: The representations (a) ${\bf 1}$ and (b) ${\bf 8}$ of
${\bf 3}\times\overline{\bf 3}$ in flavor $SU(3)$. Here $Y$ is the
hypercharge, $I$ isospin and $I_3$ the third component of the isospin.\\

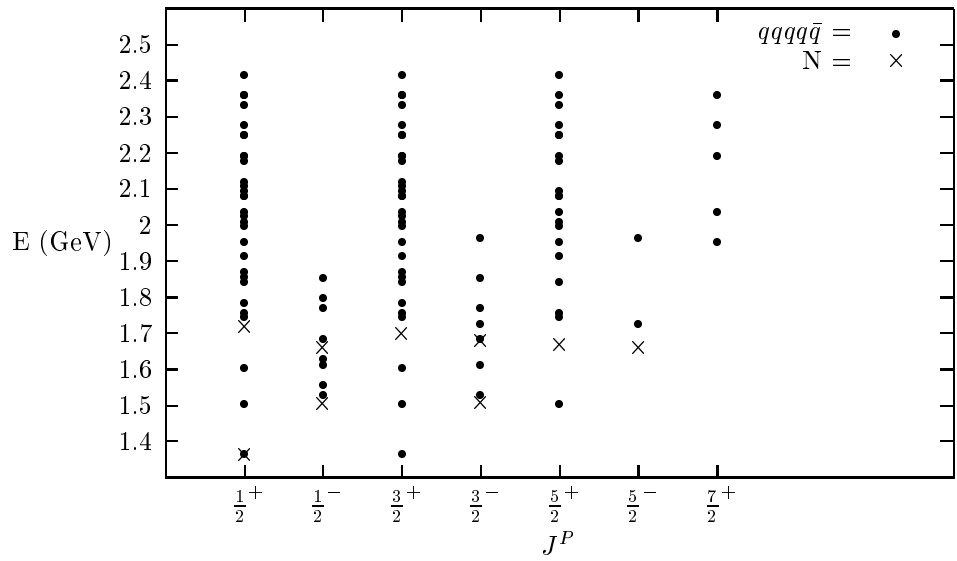
Figure 5: Empirical N states (real part of pole position)
compared to $qqqq\bar q$ states with
four-quark flavor symmetries $\lbrack 31\rbrack_F$, $\lbrack 22\rbrack_F$
and $\lbrack 211\rbrack_F$. Above E denotes the energy in GeV, $J$ total
spin and $P$ parity.\\

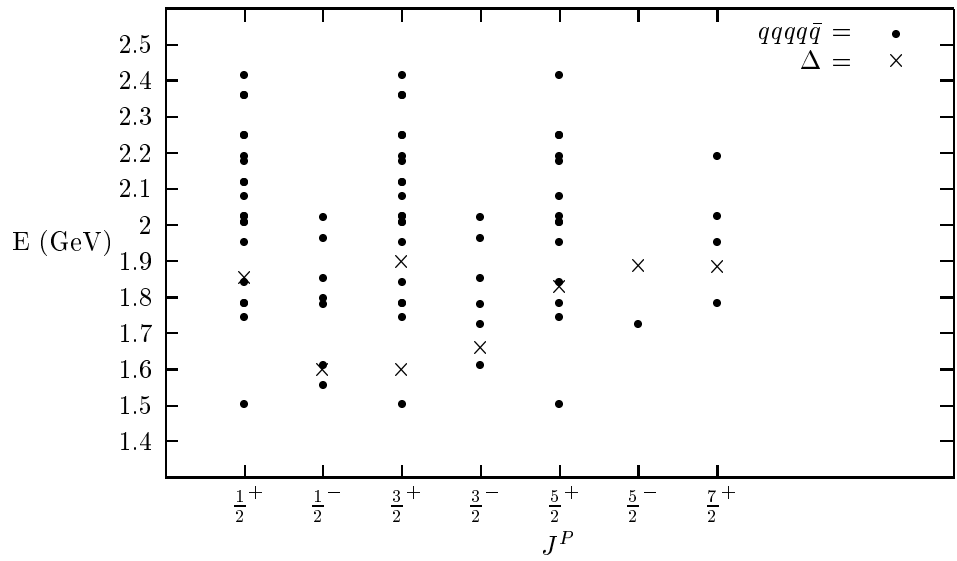
Figure 6: Empirical $\Delta$ states (real part of pole position)
compared to $qqqq\bar q$ states with
four-quark flavor symmetries $\lbrack 4\rbrack_F$ and $\lbrack 31\rbrack_F$.
Above E denotes the energy in GeV, $J$ total spin and $P$ parity.\\

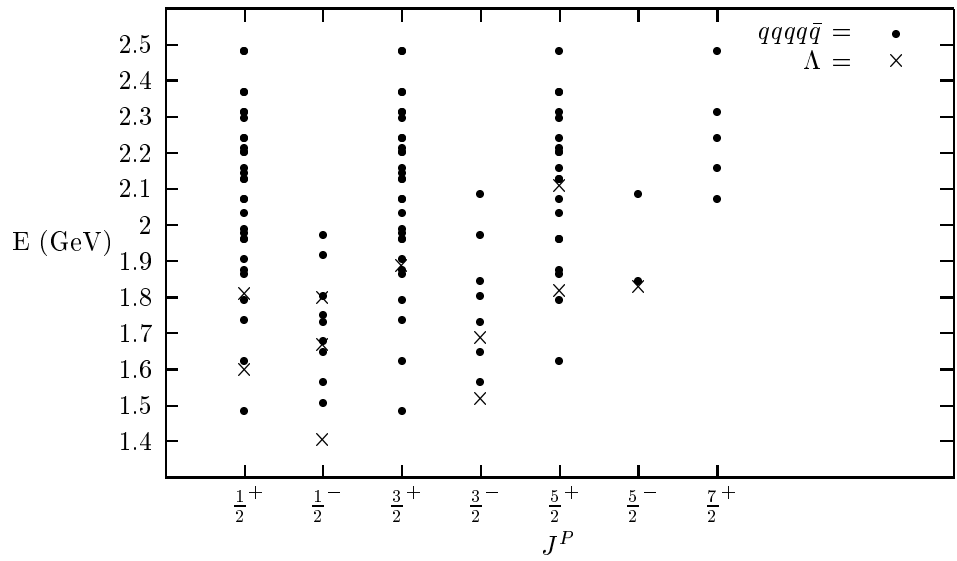
Figure 7: Empirical $\Lambda$ states compared to $qqqq\bar q$ states with
four-quark flavor symmetries $\lbrack 31\rbrack_F$, $\lbrack 22\rbrack_F$
and $\lbrack 211\rbrack_F$. Above E denotes the energy in GeV, $J$ total
spin and $P$ parity.\\

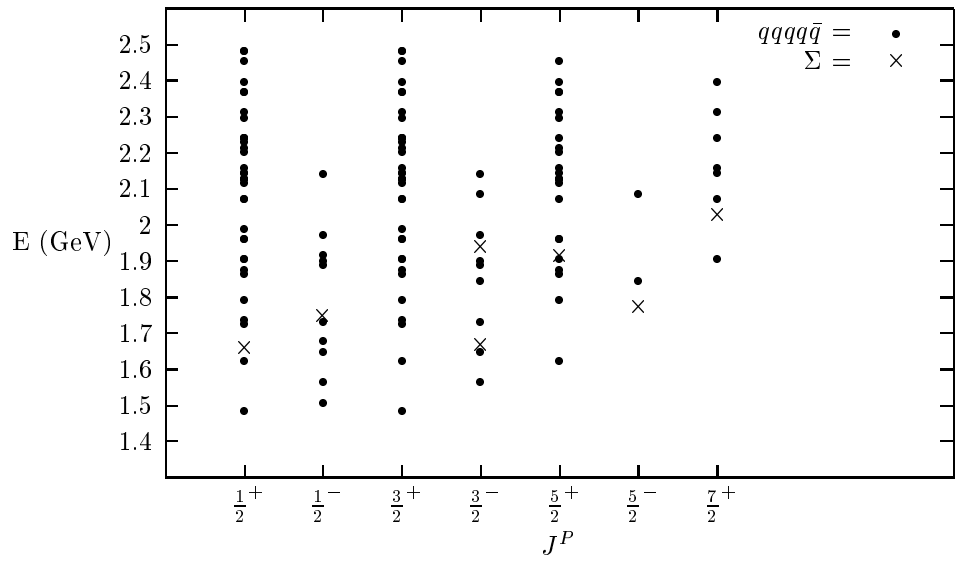
Figure 8: Empirical $\Sigma$ states compared to $qqqq\bar q$ states with
four-quark flavor symmetries $\lbrack 4\rbrack_F$, $\lbrack 31\rbrack_F$,
$\lbrack 22\rbrack_F$ and $\lbrack 211\rbrack_F$. Above E denotes the
energy in GeV, $J$ total spin and $P$ parity.\\

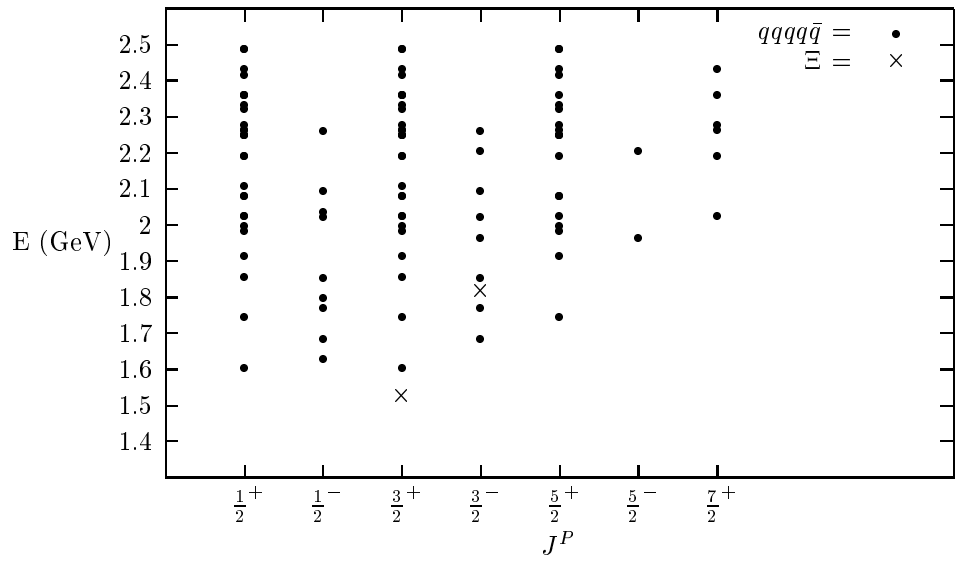
Figure 9: Empirical $\Xi$ states compared to $qqqq\bar q$ states with
four-quark flavor symmetries $\lbrack 4\rbrack_F$, $\lbrack 31\rbrack_F$,
$\lbrack 22\rbrack_F$ and $\lbrack 211\rbrack_F$. Above E denotes the
energy in GeV, $J$ total spin and $P$ parity.\\

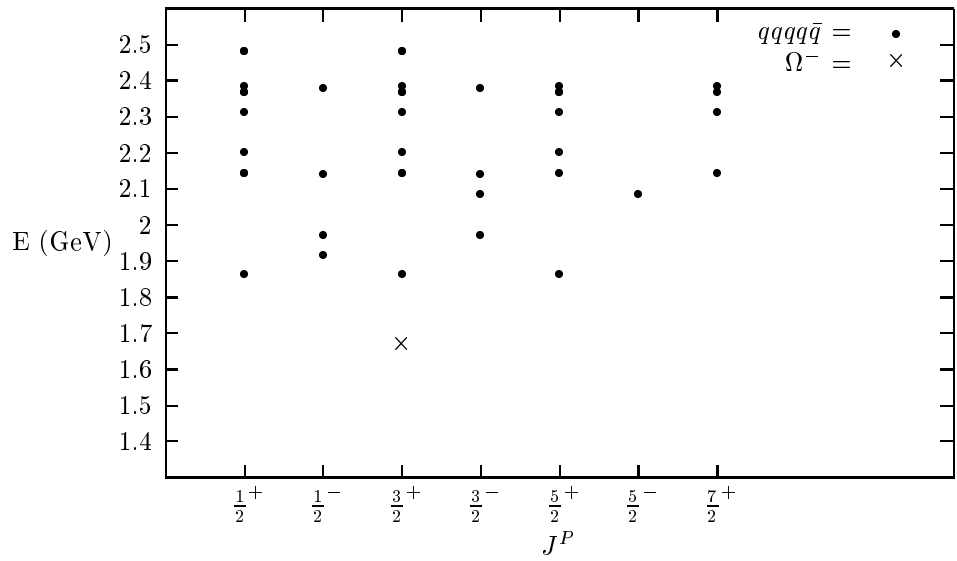
Figure 10: Empirical $\Omega^-$ state compared to $qqqq\bar q$ states with
four-quark flavor symmetries $\lbrack 4\rbrack_F$ and $\lbrack 31\rbrack_F$.
Above E denotes the energy in GeV, $J$ total spin and $P$ parity.\\


\newpage

\centerline{\bf Table 1}

\vspace{0.5cm}
\noindent Matrix elements of the squared Casimir operator $C_2^{(n)}$ for
different symmetries $\lbrack f\rbrack$ of the four-quark system.

\begin{center}
\begin{tabular}{|c|c|c|c|} \hline
&&& \\
$\lbrack f\rbrack$ & $C_2^{(6)}$ & $C_2^{(3)}$ & $C_2^{(2)}$ \\
&&& \\
\hline
&&& \\
$\lbrack 4\rbrack$ & ${50\over3}$ & ${28\over3}$ & $6$ \\
&&& \\
$\lbrack 31\rbrack$ & ${38\over3}$ & ${16\over3}$ & $2$ \\
&&& \\
$\lbrack 22\rbrack$ & ${32\over3}$ & ${10\over3}$ & $0$ \\
&&& \\
$\lbrack 211\rbrack$ & ${26\over3}$ & ${4\over3}$ & $-$ \\
&&& \\
\hline
\end{tabular}
\end{center}

\newpage

\centerline{\bf Table 2}
\vspace{0.5cm}
\noindent Matrix elements of the chiral hyperfine interaction (1.1) for the
ground state of the four-quark system with
different flavor-spin symmetries $\lbrack 31\rbrack_{FS}
\lbrack f\rbrack_F\lbrack f\rbrack_S$.

\begin{center}
\begin{tabular}{|l|r|} \hline
& \\
$\lbrack 4\rbrack_X\lbrack 1111\rbrack_{CFS}\lbrack 211\rbrack_C$
&  \\
$\lbrack f\rbrack_{FS}\lbrack f\rbrack_F\lbrack f\rbrack_S$
& $<H_{\chi}>$  \\
& \\
\hline
& \\
$\lbrack 31\rbrack_{FS}\lbrack 211\rbrack_F\lbrack 22\rbrack_S$ &
$-16\ C_{\chi}$ \\
& \\
$\lbrack 31\rbrack_{FS}\lbrack 211\rbrack_F\lbrack 31\rbrack_S$ &
$-{40\over3}\ C_{\chi}$ \\
& \\
$\lbrack 31\rbrack_{FS}\lbrack 22\rbrack_F\lbrack 31\rbrack_S$ &
$-{28\over3}\ C_{\chi}$ \\
& \\
$\lbrack 31\rbrack_{FS}\lbrack 31\rbrack_F\lbrack 22\rbrack_S$ &
$-8\ C_{\chi}$ \\
& \\
$\lbrack 31\rbrack_{FS}\lbrack 31\rbrack_F\lbrack 31\rbrack_S$ &
$-{16\over3}\ C_{\chi}$ \\
& \\
$\lbrack 31\rbrack_{FS}\lbrack 31\rbrack_F\lbrack 4\rbrack_S$ &
$0\quad$ \\
& \\
$\lbrack 31\rbrack_{FS}\lbrack 4\rbrack_F\lbrack 31\rbrack_S$ &
${8\over3}\ C_{\chi}$ \\
& \\
\hline
\end{tabular}
\end{center}

\newpage

\centerline{\bf Table 3}
\vspace{0.5cm}
\noindent Matrix elements of the chiral hyperfine interaction (1.1) for the
first excited state ($L=1$) of the four-quark system with
different flavor-spin symmetries $\lbrack f\rbrack_{FS}
\lbrack f\rbrack_F\lbrack f\rbrack_S$.

\begin{center}
\begin{tabular}{|l|r|l|r|} \hline
&&& \\
$\lbrack 31\rbrack_X\lbrack 211\rbrack_{CFS}\lbrack 211\rbrack_C$ & &
$\lbrack 31\rbrack_X\lbrack 211\rbrack_{CFS}\lbrack 211\rbrack_C$
& \\
$\lbrack f\rbrack_{FS}\lbrack f\rbrack_F\lbrack f\rbrack_S$ &
$<H_{\chi}>$ &
$\lbrack f\rbrack_{FS}\lbrack f\rbrack_F\lbrack f\rbrack_S$
& $<H_{\chi}>$ \\
&&& \\ \hline
&&& \\
$\lbrack 4\rbrack_{FS}\lbrack 22\rbrack_F\lbrack 22\rbrack_S$ &
$-28\ C_{\chi}$ &
$\lbrack 211\rbrack_{FS}\lbrack 211\rbrack_F\lbrack 22\rbrack_S$ &
$0\quad$ \\
&&& \\
$\lbrack 4\rbrack_{FS}\lbrack 31\rbrack_F\lbrack 31\rbrack_S$ &
$-{64\over3}\ C_{\chi}$ &
$\lbrack 31\rbrack_{FS}\lbrack 31\rbrack_F\lbrack 4\rbrack_S$ &
$0\quad$ \\
&&& \\
$\lbrack 31\rbrack_{FS}\lbrack 211\rbrack_F\lbrack 22\rbrack_S$ &
$-16\ C_{\chi}$ &
$\lbrack 211\rbrack_{FS}\lbrack 211\rbrack_F\lbrack 31\rbrack_S$ &
${8\over3}\ C_{\chi}$  \\
&&& \\
$\lbrack 31\rbrack_{FS}\lbrack 211\rbrack_F\lbrack 31\rbrack_S$ &
$-{40\over3}\ C_{\chi}$ &
$\lbrack 22\rbrack_{FS}\lbrack 31\rbrack_F\lbrack 31\rbrack_S$ &
${8\over3}\ C_{\chi}$ \\
&&& \\
$\lbrack 31\rbrack_{FS}\lbrack 22\rbrack_F\lbrack 31\rbrack_S$ &
$-{28\over3}\ C_{\chi}$ &
$\lbrack 31\rbrack_{FS}\lbrack 4\rbrack_F\lbrack 31\rbrack_S$ &
${8\over3}\ C_{\chi}$ \\
&&& \\
$\lbrack 31\rbrack_{FS}\lbrack 31\rbrack_F\lbrack 22\rbrack_S$ &
$-8\ C_{\chi}$ &
$\lbrack 22\rbrack_{FS}\lbrack 22\rbrack_F\lbrack 4\rbrack_S$ &
$4\ C_{\chi}$ \\
&&& \\
$\lbrack 4\rbrack_{FS}\lbrack 4\rbrack_F\lbrack 4\rbrack_S$ &
$-8\ C_{\chi}$ &
$\lbrack 211\rbrack_{FS}\lbrack 22\rbrack_F\lbrack 31\rbrack_S$ &
${20\over3}\ C_{\chi}$ \\
&&& \\
$\lbrack 22\rbrack_{FS}\lbrack 211\rbrack_F\lbrack 31\rbrack_S$ &
$-{16\over3}\ C_{\chi}$ &
$\lbrack 211\rbrack_{FS}\lbrack 211\rbrack_F\lbrack 4\rbrack_S$ &
$8\ C_{\chi}$ \\
&&& \\
$\lbrack 31\rbrack_{FS}\lbrack 31\rbrack_F\lbrack 31\rbrack_S$ &
$-{16\over3}\ C_{\chi}$ &
$\lbrack 211\rbrack_{FS}\lbrack 31\rbrack_F\lbrack 22\rbrack_S$ &
$8\ C_{\chi}$ \\
&&& \\
$\lbrack 22\rbrack_{FS}\lbrack 22\rbrack_F\lbrack 22\rbrack_S$ &
$-4\ C_{\chi}$ &
$\lbrack 22\rbrack_{FS}\lbrack 4\rbrack_F\lbrack 22\rbrack_S$ &
$8\ C_{\chi}$ \\
&&& \\
&&
$\lbrack 211\rbrack_{FS}\lbrack 31\rbrack_F\lbrack 31\rbrack_S$ &
${32\over3}\ C_{\chi}$ \\
&&& \\
\hline
\end{tabular}
\end{center}

\newpage

\centerline{\bf Table 4}
\vspace{0.5cm}

\noindent Total angular momentum $J$ and parity $P$ of the $qqqq\bar q$
system for different spin symmetries $\lbrack f\rbrack_S$ of the four-quark
subsystem. Below $S(qqqq)$ is the spin of the four-quark system,
$S(qqqq\bar q)$ the spin  and L the angular
momentum of the $qqqq\bar q$ system.

\begin{center}
\begin{tabular}{|l|r|r|r|r|}
\hline
&&&& \\
$\lbrack f\rbrack_{S(qqqq)}$ & $S(qqqq)$ & $S(qqqq\bar q)$
& $J(L=0)^P$ & $J(L=1)^P$ \\
&&&& \\
\hline
&&&& \\
$\quad\lbrack 4\rbrack_S$ & 2\quad\quad & ${5\over2}$\quad\quad
& ${5\over2}^-$\quad\quad & ${7\over2}^+$, ${5\over2}^+$, ${3\over2}^+$ \\
&&&& \\
& & ${3\over2}$\quad\quad & ${3\over2}^-$\quad\quad &
${5\over2}^+$, ${3\over2}^+$, ${1\over2}^+$ \\
&&&& \\
\hline
&&&& \\
$\quad\lbrack 31\rbrack_S$ & 1\quad\quad & ${3\over2}$\quad\quad
& ${3\over2}^-$\quad\quad & ${5\over2}^+$, ${3\over2}^+$, ${1\over2}^+$ \\
&&&& \\
&  & ${1\over2}$\quad\quad & ${1\over2}^-$\quad\quad &
${3\over2}^+$, ${1\over2}^+$ \\
&&&& \\
\hline
&&&& \\
$\quad\lbrack 22\rbrack_S$ & 0\quad\quad & ${1\over2}$\quad\quad
& ${1\over2}^-$\quad\quad & ${3\over2}^+$, ${1\over2}^+$ \\
&&&& \\
\hline
\end{tabular}
\end{center}

\newpage

\centerline{\bf Table 5}
\vspace{0.5cm}

\noindent The negative parity ($L=0$) $qqqq\bar q$ states that have the
quantum numbers of the nucleon and $\Delta$ resonance
states, which are predicted to have energies below 1900 MeV, and therefore
are candidates for mixing with the corresponding $qqq$ states. The
corresponding empirically known negative parity resonances, which are
likely to have much admixtures are also listed.

\begin{center}
\begin{tabular}{|l|c|r|r|r|} \hline
$\lbrack 4\rbrack_X\lbrack 1111\rbrack_{CFS}\lbrack 211\rbrack_C$ & Energy
 &  & & \\
$\lbrack f\rbrack_{FS}\lbrack f\rbrack_F\lbrack f\rbrack_S$
& (MeV) & $J^P$ \qquad & N (emp.) \quad\qquad & $\Delta$ (emp.)
\quad\\
\hline
& & & &\\
$\lbrack 31\rbrack_{FS}\lbrack 22\rbrack_F\lbrack 31\rbrack_S$ &
$1529$ & ${1\over2}^-$ & N(1535) & $-$\qquad\qquad \\
& & &  & \\
& & ${3\over2}^-$& N(1520) &$-$\qquad\qquad\\
& & & &\\
$\lbrack 31\rbrack_{FS}\lbrack 31\rbrack_F\lbrack 22\rbrack_S$ &
$1557$ &  ${1\over2}^-$ & N(1535) & $\Delta(1620)$ \\
& & & & \\
$\lbrack 31\rbrack_{FS}\lbrack 31\rbrack_F\lbrack 31\rbrack_S$ &
$1613$ & ${1\over2}^-$ & N(1650)& $\Delta(1620)$ \\
& & & & \\
& & ${3\over2}^-$ & N(1700)& $\Delta(1700)$\\
& & & &\\
$\lbrack 31\rbrack_{FS}\lbrack 211\rbrack_F\lbrack 22\rbrack_S$ ($qqqs\bar s$)&
$1629$ & ${1\over2}^-$ & N(1650) & $-$\qquad\qquad \\
&  & & &\\
$\lbrack 31\rbrack_{FS}\lbrack 211\rbrack_F\lbrack 31\rbrack_S$ ($qqqs\bar s$)&
$1685$ &${1\over2}^-$ & N(1650)
 & $-$\qquad\qquad \\
& & & &\\
& &${3\over2}^-$ & N(1700) &$-$\qquad\qquad\\
& & & & \\
$\lbrack 31\rbrack_{FS}\lbrack 31\rbrack_F\lbrack 4\rbrack_S$ &
$1725$ &${3\over2}^-$ & N(1700)
 & $\Delta(1700)$ \\
& & & &\\
& &${5\over2}^-$ & N(1675) & \\
& & & & \\
$\lbrack 31\rbrack_{FS}\lbrack 22\rbrack_F\lbrack 31\rbrack_S$ ($qqqs\bar s$)&
$1769$ &${1\over2}^-$ &
 & $-$\qquad\qquad \\
& & & &\\
& &${3\over2}^-$ & N(1700) &$-$\qquad\qquad\\
& & & & \\
$\lbrack 31\rbrack_{FS}\lbrack 4\rbrack_F\lbrack 31\rbrack_S$
& $1781$ &${1\over2}^-$ &
 $-$\qquad\qquad & $\Delta(1900)$(**)\\
& & & &\\
& &${3\over2}^-$ & $-$\qquad\qquad & $\Delta(1700)$\\
& & & & \\
$\lbrack 31\rbrack_{FS}\lbrack 31\rbrack_F\lbrack 22\rbrack_S$ ($qqqs\bar s$)&
$1797$ &${1\over2}^-$ &
 & $\Delta(1900)$(**)\\
& & & &\\
$\lbrack 31\rbrack_{FS}\lbrack 31\rbrack_F\lbrack 31\rbrack_S$ ($qqqs\bar s$)&
$1853$ &${1\over2}^-$ &
 & $\Delta(1900)$(**) \\
& & & &\\
& &${3\over2}^-$ &  & \\
& & & & \\
\hline
\end{tabular}
\end{center}

\newpage

\centerline{\bf Table 6}
\vspace{0.5cm}

\noindent The positive parity ($L=1$) states that have the
quantum numbers of the nucleon and the
$\Delta$ resonance states, which are predicted to have energies below
1900 MeV, and therefore are candidates for mixing with the corresponding
$qqq$ states. The corresponding empirically known positive parity
resonances, which are likely to have much admixtures are also listed.
The corresponding predicted $qqq$ states in this energy range, which have
not yet been identified empirically, are also shown, with a question mark.
The symbol $+$ indicates that there is a corresponding state with the
antiquark in the p-state and the four quarks in a state with symmetry
$\lbrack 4\rbrack_X\lbrack 1111\rbrack_{CFS}\lbrack 211\rbrack_C
\lbrack 31\rbrack_{FS}\lbrack f\rbrack_{F}\lbrack f\rbrack_S$
at the indicated energy.

\begin{center}
\begin{tabular}{|l|c|r|r|r|} \hline
$\lbrack 31\rbrack_X\lbrack 211\rbrack_{CFS}\lbrack 211\rbrack_C$
& Energy &  \qquad & & \\
$\lbrack f\rbrack_{FS}\lbrack f\rbrack_F\lbrack f\rbrack_S$
& (MeV) & $J^P$ \ \qquad & N (emp.)\quad & $\Delta$ (emp.)
\quad\\
\hline
& & & & \\
$\lbrack 4\rbrack_{FS}\lbrack 22\rbrack_F\lbrack 22\rbrack_S$ &
1365 & ${1\over2}^+$ & N(1440) &
$-$ \qquad\qquad\\
&  & & & \\
& & ${3\over2}^+$& & $-$ \qquad\qquad\\
& & & & \\
$\lbrack 4\rbrack_{FS}\lbrack 31\rbrack_F\lbrack 31\rbrack_S$ &
$1505$ & ${1\over2}^+$ & N(1440) &   \\
&  & & & \\
& & ${3\over2}^+$& & $\Delta(1600)$ \\
& & & & \\
& & ${5\over2}^+$&  &  \\
& & & & \\
$\lbrack 4\rbrack_{FS}\lbrack 22\rbrack_F\lbrack 22\rbrack_S$ ($qqqs\bar s$)&
1605 & ${1\over2}^+$ & N(1710) &
$-$ \qquad\qquad\\
&  & & & \\
& & ${3\over2}^+$& N(1720)  &
$-$ \qquad\qquad  \\
& & & & \\
$\lbrack 4\rbrack_{FS}\lbrack 31\rbrack_F\lbrack 31\rbrack_S$ ($qqqs\bar s$)&
$1745$ & ${1\over2}^+$ & N(1710) & $\Delta(1750)$(*) \\
&  & & & \\
& & ${3\over2}^+$& N(1720) & $\Delta(?)$, $\Delta(1600)$ \\
& & & & \\
& & ${5\over2}^+$& N(1680) & $\Delta(?)$, $\Delta(1905)$ \\
& & & & \\
$\lbrack 31\rbrack_{FS}\lbrack 22\rbrack_F\lbrack 31\rbrack_S$ &
$1757$ & ${1\over2}^+$ & N(1710) & $-$ \qquad\qquad  \\
& $+$ &  & & \\
& & ${3\over2}^+$& N(1720) & $-$ \qquad\qquad \\
& & & & \\
& & ${5\over2}^+$& N(1680)& $-$ \qquad\qquad \\
& & & & \\
\hline
\end{tabular}
\end{center}

\newpage

\centerline{\bf Table 6, continued}
\vspace{0.5cm}

\begin{center}
\begin{tabular}{|l|c|r|r|r|} \hline
$\lbrack 31\rbrack_X\lbrack 211\rbrack_{CFS}\lbrack 211\rbrack_C$
& Energy &  \qquad & & \\
$\lbrack f\rbrack_{FS}\lbrack f\rbrack_F\lbrack f\rbrack_S$
& (MeV) & $J^P$ \ \qquad & N (emp.)\quad & $\Delta$ (emp.)
\quad\\
\hline
& & & & \\
$\lbrack 31\rbrack_{FS}\lbrack 31\rbrack_F\lbrack 22\rbrack_S$ &
$1785$ & ${1\over2}^+$ & N(1710) & $\Delta(1750)$(*)\\
& $+$ & & & \\
& & ${3\over2}^+$& N(1720) &  \\
& & & & \\
$\lbrack 4\rbrack_{FS}\lbrack 4\rbrack_F\lbrack 4\rbrack_S$ &
$1785$ & ${1\over2}^+$ & $-$ \quad\quad & $\Delta(1750)$(*)\\
&  & & & \\
& & ${3\over2}^+$& $-$ \quad\quad & \\
& & & & \\
& & ${5\over2}^+$& $-$ \quad\quad & $\Delta(1905)$\\
& & & & \\
& & ${7\over2}^+$& $-$ \quad\quad &\\
& & & & \\
$\lbrack 31\rbrack_{FS}\lbrack 31\rbrack_F\lbrack 31\rbrack_S$ &
$1841$ & ${1\over2}^+$ & N(?)&  $\Delta(1910)$\\
& $+$ & & & \\
& & ${3\over2}^+$ & N(?) & $\Delta(1920)$\\
& & & & \\
& & ${5\over2}^+$ & N(?) & $\Delta(1905)$\\
& & & & \\
$\lbrack 31\rbrack_{FS}\lbrack 211\rbrack_F\lbrack 22\rbrack_S$ ($qqqs\bar s$)&
$1857$ & ${1\over2}^+$ & N(?) &  $-$ \qquad\qquad \\
& $+$ & & & \\
& & ${3\over3}^+$ & N(?) & $-$ \qquad\qquad \\
& & & & \\
$\lbrack 22\rbrack_{FS}\lbrack 22\rbrack_F\lbrack 22\rbrack_S$ &
$1869$ & ${1\over2}^+$ & N(?)  &  $-$ \qquad\qquad \\
& $+$ & & & \\
& & ${3\over3}^+$ & N(?) & $-$ \qquad\qquad \\
& & & & \\
\hline
\end{tabular}
\end{center}

\newpage

\centerline{\bf Table 7}
\vspace{0.5cm}

\noindent The negative parity ($L=0$) $qqqq\bar q$ states that have the
quantum numbers of the $\Lambda$ and $\Sigma$ hyperon resonance
states, which are predicted to have energies up to 2000 MeV, and therefore
are candidates for mixing with the corresponding $qqq$ states. The
corresponding empirically known negative parity resonances, which are
likely to have much admixtures are also listed.
The corresponding predicted $qqq$ states in this energy range, which have
not yet been identified empirically, are also shown, with a question mark.

\begin{center}
\begin{tabular}{|l|c|r|r|r|} \hline
$\lbrack 4\rbrack_X\lbrack 1111\rbrack_{CFS}\lbrack 211\rbrack_C$
& Energy &  & &\\
$\lbrack f\rbrack_{FS}\lbrack f\rbrack_F\lbrack f\rbrack_S$
& (MeV) & $J^P$ \qquad & $\Lambda$ (emp.)\quad\qquad
& $\Sigma$ (emp.)\quad
\\
\hline
& & & &\\
$\lbrack 31\rbrack_{FS}\lbrack 211\rbrack_F\lbrack 22\rbrack_S$ &
$1509$ & ${1\over2}^-$ & $\Lambda(1405)$
& $\Sigma(1560)$(**)  \\
& & & & \\
$\lbrack 31\rbrack_{FS}\lbrack 211\rbrack_F\lbrack 31\rbrack_S$ &
$1565$ & ${1\over2}^-$ &  & $\Sigma(1560)$(**),
$\Sigma(1620)$(**)\\
& & & & \\
& & ${3\over2}^-$ & $\Lambda(1520)$ &
$\Sigma(1580)$(**) \\
& & & & \\
$\lbrack 31\rbrack_{FS}\lbrack 22\rbrack_F\lbrack 31\rbrack_S$ &
$1649$ & ${1\over2}^-$ & $\Lambda(1670)$ & $\Sigma(1620)$(**) \\
& & & & \\
& & ${3\over2}^-$& $\Lambda(1690)$ & $\Sigma(1670)$ \\
& & & & \\
$\lbrack 31\rbrack_{FS}\lbrack 31\rbrack_F\lbrack 22\rbrack_S$ &
$1677$ & ${1\over2}^-$ & $\Lambda(1670)$ & $\Sigma(1690)$(**)\\
& & & & \\
$\lbrack 31\rbrack_{FS}\lbrack 31\rbrack_F\lbrack 31\rbrack_S$ &
$1733$ & ${1\over2}^-$ & $\Lambda(1800)$ & $\Sigma(1750)$ \\
& & & & \\
& & ${3\over2}^-$&  $\Lambda(1690)$& $\Sigma(1670)$, $\Sigma(?)$ \\
& & & & \\
$\lbrack 31\rbrack_{FS}\lbrack 211\rbrack_F\lbrack 22\rbrack_S$ ($qqss\bar s$)&
$1749$ & ${1\over2}^-$ & $\Lambda(1800)$ & $-$ \qquad\qquad \\
& & & & \\
$\lbrack 31\rbrack_{FS}\lbrack 211\rbrack_F\lbrack 31\rbrack_S$ ($qqss\bar s$)&
$1805$ & ${1\over2}^-$ & $\Lambda(1800)$ & $-$ \qquad\qquad\\
& & & & \\
& & ${3\over2}^-$ & $\Lambda(?)$ &$-$ \qquad\qquad\\
& & & & \\
\hline
\end{tabular}
\end{center}

\newpage

\centerline{\bf Table 7, continued}
\vspace{0.5cm}

\begin{center}
\begin{tabular}{|l|c|r|r|r|} \hline
$\lbrack 4\rbrack_X\lbrack 1111\rbrack_{CFS}\lbrack 211\rbrack_C$
& Energy &  & &\\
$\lbrack f\rbrack_{FS}\lbrack f\rbrack_F\lbrack f\rbrack_S$
& (MeV) & $J^P$ \qquad & $\Lambda$ (emp.)\quad\qquad
& $\Sigma$ (emp.)\quad
\\
\hline
& & & &\\
$\lbrack 31\rbrack_{FS}\lbrack 31\rbrack_F\lbrack 4\rbrack_S$&
$1845$ & ${3\over2}^-$ &  & $\Sigma(1940)$\\
& & & & \\
& & ${5\over2}^-$ & $\Lambda(1830)$ & $\Sigma(1775)$\\
& & & & \\
$\lbrack 31\rbrack_{FS}\lbrack 22\rbrack_F\lbrack 31\rbrack_S$ ($qqss\bar s$)&
$1889$ & ${1\over2}^-$ & $-$ \qquad\qquad & \\
& & & & \\
& & ${3\over2}^-$ & $-$ \qquad\qquad & $\Sigma(1940)$\\
& & & & \\
$\lbrack 31\rbrack_{FS}\lbrack 4\rbrack_F\lbrack 31\rbrack_S$ &
$1901$ & ${1\over2}^-$ & $-$ \qquad\qquad & \\
& & & & \\
& & ${3\over2}^-$ & $-$ \qquad\qquad & $\Sigma(1940)$\\
& & & & \\
$\lbrack 31\rbrack_{FS}\lbrack 31\rbrack_F\lbrack 22\rbrack_S$ ($qqss\bar s$)&
$1917$ & ${1\over2}^-$ & & \\
& & & & \\
$\lbrack 31\rbrack_{FS}\lbrack 31\rbrack_F\lbrack 31\rbrack_S$ ($qqss\bar s$)&
$1973$ & ${1\over2}^-$ & & \\
& & & & \\
& & ${3\over2}^-$ & & $\Sigma(1940)$\\
& & & & \\
\hline
\end{tabular}
\end{center}

\newpage

\centerline{\bf Table 8}
\vspace{0.5cm}

\noindent The positive parity ($L=1$) states that have the
quantum numbers of the $\Lambda$ and the
$\Sigma$ hyperon resonance states, which are predicted to have energies up
to 2000 MeV, and therefore are candidates for mixing with the corresponding
$qqq$ states. The corresponding empirically known positive parity
resonances, which are likely to have much admixtures are also listed.
The corresponding predicted $qqq$ states in this energy range, which have
not yet been identified empirically, are also shown, with a question mark.
The symbol $+$ indicates that there is a corresponding state with the
antiquark in the p-state and the four quarks in a state with symmetry
$\lbrack 4\rbrack_X\lbrack 1111\rbrack_{CFS}\lbrack 211\rbrack_C
\lbrack 31\rbrack_{FS}\lbrack f\rbrack_{F}\lbrack f\rbrack_S$
at the indicated energy.

\begin{center}
\begin{tabular}{|l|c|r|r|r|} \hline
$\lbrack 31\rbrack_X\lbrack 211\rbrack_{CFS}\lbrack 211\rbrack_C$
& Energy &  \qquad & &\\
$\lbrack f\rbrack_{FS}\lbrack f\rbrack_F\lbrack f\rbrack_S$
& (MeV) & $J^P$ \ \qquad & $\Lambda$ (emp.)\quad
& $\Sigma$ (emp.)\quad\\
\hline
& & & & \\
$\lbrack 4\rbrack_{FS}\lbrack 22\rbrack_F\lbrack 22\rbrack_S$ &
$1485$ & ${1\over2}^+$ & $\Lambda(1600)$ & $\Sigma(1480)$(*)?, $\Sigma(1600)$ \\
&  & & & \\
& & ${3\over2}^+$& & $\Sigma(1480)$(*)?\\
& & & &\\
$\lbrack 4\rbrack_{FS}\lbrack 31\rbrack_F\lbrack 31\rbrack_S$ &
$1625$ & ${1\over2}^+$ & $\Lambda(1600)$ &
$\Sigma(1660)$\\
& & & & \\
& & ${3\over2}^+$ &  & \\
& & & &\\
& & ${5\over2}^+$ & &\\
& & & &\\
$\lbrack 4\rbrack_{FS}\lbrack 22\rbrack_F\lbrack 22\rbrack_S$ ($qqss\bar s$)&
$1725$ & ${1\over2}^+$ & $-$\qquad\qquad & $\Sigma(1660)$, $\Sigma(1770)(*)$ \\
&  & & & \\
& & ${3\over2}^+$& $-$\qquad\qquad & $\Sigma(?)$\\
& & & &\\
$\lbrack 31\rbrack_{FS}\lbrack 211\rbrack_F\lbrack 22\rbrack_S$ &
$1737$ & ${1\over2}^+$ & $\Lambda(1810)$ & $\Sigma(1660)$, $\Sigma(1770)$(*)\\
& $+$ & & & \\
& & ${3\over2}^+$ & $\Lambda(1890)$ & $\Sigma(?)$ \\
& & & &\\
$\lbrack 31\rbrack_{FS}\lbrack 211\rbrack_F\lbrack 31\rbrack_S$ &
$1793$ & ${1\over2}^+$ & $\Lambda(1810)$
& $\Sigma(1770)$(**) \\
& $+$ & & &\\
& & ${3\over2}^+$& $\Lambda(1890)$
& $\Sigma(?)$, $\Sigma(1840)$(*)\\
& & & &\\
& & ${5\over2}^+$& $\Lambda(1820)$ & \\
& & & & \\
$\lbrack 4\rbrack_{FS}\lbrack 31\rbrack_F\lbrack 31\rbrack_S$ ($qqss\bar s$)&
$1865$ & ${1\over2}^+$ & $\Lambda(1810)$ &
$\Sigma(1880)$(**)\\
& & & & \\
& & ${3\over2}^+$ & $\Lambda(1890)$ & $\Sigma(1840)$(*) \\
& & & &\\
& & ${5\over2}^+$ & $\Lambda(1820)$ & $\Sigma(?)$,
$\Sigma(1915)$\\
& & & &\\
$\lbrack 31\rbrack_{FS}\lbrack 22\rbrack_F\lbrack 31\rbrack_S$ &
$1877$ & ${1\over2}^+$ & $\Lambda(1810)$ & $\Sigma(1880)$(**)\\
& $+$ & & & \\
& & ${3\over2}^+$ & & $\Sigma(1840)$(*), $\Sigma(?)$\\
& & & & \\
& & ${5\over2}^+$ &  & $\Sigma(1915)$\\
& & & &\\
\hline
\end{tabular}
\end{center}

\newpage

\centerline{\bf Table 8, continued}

\begin{center}
\begin{tabular}{|l|c|r|r|r|} \hline
$\lbrack 31\rbrack_X\lbrack 211\rbrack_{CFS}\lbrack 211\rbrack_C$
& Energy &  \qquad & &\\
$\lbrack f\rbrack_{FS}\lbrack f\rbrack_F\lbrack f\rbrack_S$
& (MeV) & $J^P$ \ \qquad & $\Lambda$ (emp.)\quad
& $\Sigma$ (emp.)\quad\\
\hline
& & & &\\
$\lbrack 31\rbrack_{FS}\lbrack 31\rbrack_F\lbrack 22\rbrack_S$ &
$1905$ & ${1\over2}^+$ & $\Lambda(1810)$ & $\Sigma(1880)$(**), $\Sigma(?)$\\
& $+$ & & & \\
& & ${3\over2}^+$ & $\Lambda(?)$ & $\Sigma(?)$\\
& & & &\\
$\lbrack 4\rbrack_{FS}\lbrack 4\rbrack_F\lbrack 4\rbrack_S$ &
$1905$ & ${1\over2}^+$ & $-$\quad\quad & $\Sigma(1880)$(**), $\Sigma(?)$\\
& & & & \\
& & ${3\over2}^+$ & $-$\quad\quad & $\Sigma(?)$\\
& & & &\\
& & ${5\over2}^+$ & $-$\quad\quad & $\Sigma(1915)$\\
& & & &\\
& & ${7\over2}^+$ & $-$\quad\quad & \\
& & & &\\
$\lbrack 22\rbrack_{FS}\lbrack 211\rbrack_F\lbrack 31\rbrack_S$ &
$1961$ & ${1\over2}^+$ & $\Lambda(?)$ & $\Sigma(?)$\\
& & & & \\
& & ${3\over2}^+$ & $\Lambda(?)$ & \\
& & & &\\
& & ${5\over2}^+$ & $\Lambda(?)$& $\Sigma(1915)$\\
& & & &\\
$\lbrack 31\rbrack_{FS}\lbrack 31\rbrack_F\lbrack 31\rbrack_S$ &
$1961$ & ${1\over2}^+$ & $\Lambda(?)$ & $\Sigma(?)$\\
& $+$ & & & \\
& & ${3\over2}^+$ & $\Lambda(?)$ & \\
& & & &\\
& & ${5\over2}^+$ & $\Lambda(?)$ & $\Sigma(1915)$\\
& & & &\\
$\lbrack 31\rbrack_{FS}\lbrack 211\rbrack_F\lbrack 22\rbrack_S$ ($qqss\bar s$)&
$1977$ & ${1\over2}^+$ & $\Lambda(?)$ & $-$\quad\quad  \\
& & & & \\
& & ${3\over2}^+$ & $\Lambda(?)$ & $-$\quad\quad \\
& & & &\\
$\lbrack 22\rbrack_{FS}\lbrack 22\rbrack_F\lbrack 22\rbrack_S$ &
$1989$ & ${1\over2}^+$ & $\Lambda(?)$ & $\Sigma(?)$  \\
& & & & \\
& & ${3\over2}^+$ & $\Lambda(?)$ & $\Sigma(?)$ \\
\hline
\end{tabular}
\end{center}

\newpage

\centerline{\bf Table 9}
\vspace{0.5cm}

\noindent The negative parity ($L=0$) $qqqq\bar q$ states that have the
quantum numbers of the $\Xi$ hyperon resonance
states, which are predicted to have energies up to 2100 MeV, and therefore
are candidates for mixing with the corresponding $qqq$ states.
The corresponding predicted $qqq$ states in this energy range, which have
not yet been identified empirically, are also shown, with a question mark.

\begin{center}
\begin{tabular}{|l|c|r|r|} \hline
$\lbrack 4\rbrack_X\lbrack 1111\rbrack_{CFS}\lbrack 211\rbrack_C$
& Energy & & \\
$\lbrack f\rbrack_{FS}\lbrack f\rbrack_F\lbrack f\rbrack_S$
& (MeV) & $J^P$ \qquad & $\Xi$ (emp.)\\
\hline
& & &\\
$\lbrack 31\rbrack_{FS}\lbrack 211\rbrack_F\lbrack 22\rbrack_S$ &
$1629$ & ${1\over2}^-$ & $\Xi(1690)$?\\
& & &\\
$\lbrack 31\rbrack_{FS}\lbrack 211\rbrack_F\lbrack 31\rbrack_S$ &
$1685$ & ${1\over2}^-$ & $\Xi(1690)$?\\
& & &\\
& & ${3\over2}^-$& $\Xi(1690)$?\\
& & &\\
$\lbrack 31\rbrack_{FS}\lbrack 22\rbrack_F\lbrack 31\rbrack_S$ &
$1769$ & ${1\over2}^-$ &  $\Xi(?)$\\
& & &\\
& & ${3\over2}^-$ &  $\Xi(?)$, $\Xi(1820)$\\
& & &\\
$\lbrack 31\rbrack_{FS}\lbrack 31\rbrack_F\lbrack 22\rbrack_S$ &
$1797$ & ${1\over2}^-$ & \\
& & &\\
$\lbrack 31\rbrack_{FS}\lbrack 31\rbrack_F\lbrack 31\rbrack_S$ &
$1853$ & ${1\over2}^-$ & $\Xi(?)$\\
& & &\\
& & ${3\over2}^-$&  $\Xi(1820)$,  $\Xi(?)$\\
& & &\\
$\lbrack 31\rbrack_{FS}\lbrack 31\rbrack_F\lbrack 4\rbrack_S$ &
$1965$ & ${3\over2}^-$ &  $\Xi(1950)$?\\
& & &\\
& & ${5\over2}^-$& $\Xi(1950)$?\\
& & &\\
$\lbrack 31\rbrack_{FS}\lbrack 4\rbrack_F\lbrack 31\rbrack_S$ &
$2021$ & ${1\over2}^-$ & \\
& & &\\
& & ${3\over2}^-$&\\
& & &\\
$\lbrack 31\rbrack_{FS}\lbrack 31\rbrack_F\lbrack 22\rbrack_S$ ($qsss\bar s$)&
$2037$ & ${1\over2}^-$ & \\
& & &\\
$\lbrack 31\rbrack_{FS}\lbrack 31\rbrack_F\lbrack 31\rbrack_S$ ($qsss\bar s$)&
$2093$ & ${1\over2}^-$ & \\
& & &\\
& & ${3\over2}^-$&\\
& & &\\
\hline
\end{tabular}
\end{center}

\newpage

\centerline{\bf Table 10}
\vspace{0.5cm}

\noindent The positive parity ($L=1$) states that have the
quantum numbers of the $\Xi$ hyperon resonance states,
which are predicted to have energies up
to 2100 MeV, and therefore are candidates for mixing with the corresponding
$qqq$ states.
The corresponding predicted $qqq$ states in this energy range, which have
not yet been identified empirically, are also shown, with a question mark.
The symbol $+$ indicates that there is a corresponding state with the
antiquark in the p-state and the four quarks in a state with symmetry
$\lbrack 4\rbrack_X\lbrack 1111\rbrack_{CFS}\lbrack 211\rbrack_C
\lbrack 31\rbrack_{FS}\lbrack f\rbrack_{F}\lbrack f\rbrack_S$
at the indicated energy.

\begin{center}
\begin{tabular}{|l|c|r|r|} \hline
$\lbrack 31\rbrack_X\lbrack 211\rbrack_{CFS}\lbrack 211\rbrack_C$
& Energy &  \qquad & \\
$\lbrack f\rbrack_{FS}\lbrack f\rbrack_F\lbrack f\rbrack_S$
& (MeV) & $J^P$ \ \qquad & $\Xi$ (emp.)\quad\\
\hline
& & & \\
$\lbrack 4\rbrack_{FS}\lbrack 22\rbrack_F\lbrack 22\rbrack_S$ &
$1605$ & ${1\over2}^+$ & $\Xi(1620)$(*)? \\
&  & & \\
& & ${3\over2}^+$ & $\Xi(1620)$(*)?\\
& & & \\
$\lbrack 4\rbrack_{FS}\lbrack 31\rbrack_F\lbrack 31\rbrack_S$ &
$1745$ & ${1\over2}^+$ & $\Xi(?)$\\
& & & \\
& & ${3\over2}^+$ & \\
& & & \\
& & ${5\over2}^+$ & \\
& & & \\
$\lbrack 31\rbrack_{FS}\lbrack 211\rbrack_F\lbrack 22\rbrack_S$ &
$1857$ & ${1\over2}^+$ & \\
& $+$ & & \\
& & ${3\over2}^+$ & $\Xi(?)$\\
& & & \\
$\lbrack 31\rbrack_{FS}\lbrack 211\rbrack_F\lbrack 31\rbrack_S$ &
$1913$ & ${1\over2}^+$ & $\Xi(?)$\\
& $+$  & & \\
& & ${3\over2}^+$ & $\Xi(?)$\\
& & & \\
& & ${5\over2}^+$ & $\Xi(?)$\\
& & & \\
$\lbrack 4\rbrack_{FS}\lbrack 31\rbrack_F\lbrack 31\rbrack_S$ ($qsss\bar s$)&
$1985$ & ${1\over2}^+$ & $\Xi(?)$\\
& & & \\
& & ${3\over2}^+$ & \\
& & & \\
& & ${5\over2}^+$ & \\
& & & \\
\hline
\end{tabular}
\end{center}

\newpage

\centerline{\bf Table 10, continued}

\begin{center}
\begin{tabular}{|l|c|r|r|} \hline
$\lbrack 31\rbrack_X\lbrack 211\rbrack_{CFS}\lbrack 211\rbrack_C$
& Energy &  \qquad & \\
$\lbrack f\rbrack_{FS}\lbrack f\rbrack_F\lbrack f\rbrack_S$
& (MeV) & $J^P$ \ \qquad & $\Xi$ (emp.)\quad\\
\hline
& & & \\
$\lbrack 31\rbrack_{FS}\lbrack 22\rbrack_F\lbrack 31\rbrack_S$ &
$1997$ & ${1\over2}^+$ & \\
& $+$ & & \\
& & ${3\over2}^+$ & \\
& & & \\
& & ${5\over2}^+$ & \\
& & & \\
$\lbrack 31\rbrack_{FS}\lbrack 31\rbrack_F\lbrack 22\rbrack_S$ &
$2025$ & ${1\over2}^+$ & $\Xi(?)$ \\
& $+$ & & \\
& & ${3\over2}^+$ & $\Xi(?)$ \\
& & & \\
$\lbrack 4\rbrack_{FS}\lbrack 4\rbrack_F\lbrack 4\rbrack_S$ &
$2025$ & ${1\over2}^+$ &$\Xi(?)$ \\
& & & \\
& & ${3\over2}^+$ & $\Xi(?)$ \\
& & & \\
& & ${5\over2}^+$ & $\Xi(2030)$? \\
& & & \\
& & ${7\over2}^+$ & $\Xi(2030)$?, $\Xi(?)$ \\
& & & \\
$\lbrack 22\rbrack_{FS}\lbrack 211\rbrack_F\lbrack 31\rbrack_S$ &
$2081$ & ${1\over2}^+$ & $\Xi(?)$ \\
& & & \\
& & ${3\over2}^+$ & $\Xi(?)$ \\
& & & \\
& & ${5\over2}^+$ & $\Xi(?)$ \\
& & & \\
$\lbrack 31\rbrack_{FS}\lbrack 31\rbrack_F\lbrack 31\rbrack_S$ &
$2081$ & ${1\over2}^+$ & $\Xi(?)$ \\
& $+$ & & \\
& & ${3\over2}^+$ & $\Xi(?)$ \\
& & & \\
& & ${5\over2}^+$ & $\Xi(?)$ \\
& & & \\
\hline
\end{tabular}
\end{center}

\newpage

\centerline{\bf Table 11}
\vspace{0.5cm}

\noindent The negative parity ($L=0$) $qqqq\bar q$ states that have the
quantum numbers of the $\Omega^-$ hyperon resonance
states, which are predicted to have energies up to 2400 MeV, and therefore
are candidates for mixing with the corresponding $qqq$ states.
The corresponding predicted $qqq$ states in this energy range, which have
not yet been identified empirically, are also shown, with a question mark.

\begin{center}
\begin{tabular}{|l|c|r|r|} \hline
$\lbrack 4\rbrack_X\lbrack 1111\rbrack_{CFS}\lbrack 211\rbrack_C$
& Energy & & \\
$\lbrack f\rbrack_{FS}\lbrack f\rbrack_F\lbrack f\rbrack_S$
& (MeV) & $J^P$ \qquad & $\Omega^-$ (emp.)\\
\hline
& & &\\
$\lbrack 31\rbrack_{FS}\lbrack 31\rbrack_F\lbrack 22\rbrack_S$ &
$1917$ & ${1\over2}^-$ & \\
& & &\\
$\lbrack 31\rbrack_{FS}\lbrack 31\rbrack_F\lbrack 31\rbrack_S$ &
$1973$ & ${1\over2}^-$ & $\Omega^-(?)$\\
& & &\\
& & ${3\over2}^-$&  $\Omega^-(?)$\\
& & &\\
$\lbrack 31\rbrack_{FS}\lbrack 31\rbrack_F\lbrack 4\rbrack_S$ &
$2085$ & ${3\over2}^-$ & \\
& & &\\
& & ${5\over2}^-$& \\
& & &\\
$\lbrack 31\rbrack_{FS}\lbrack 4\rbrack_F\lbrack 31\rbrack_S$ &
$2141$ & ${1\over2}^-$ &  \\
& & &\\
& & ${3\over2}^-$& \\
& & &\\
$\lbrack 31\rbrack_{FS}\lbrack 4\rbrack_F\lbrack 31\rbrack_S$ ($ssss\bar s$) &
$2381$ & ${1\over2}^-$ & $\Omega^-(2380)$(**)? \\
& & &\\
& & ${3\over2}^-$& $\Omega^-(2380)$(**)?\\
& & &\\
\hline
\end{tabular}
\end{center}

\newpage

\centerline{\bf Table 12}
\vspace{0.5cm}

\noindent The positive parity ($L=1$) states that have the
quantum numbers of the $\Omega^-$ hyperon resonance states,
which are predicted to have energies up
to 2400 MeV, and therefore are candidates for mixing with the corresponding
$qqq$ states.
The corresponding predicted $qqq$ states in this energy range, which have
not yet been identified empirically, are also shown, with a question mark.
The symbol $+$ indicates that there is a corresponding state with the
antiquark in the p-state and the four quarks in a state with symmetry
$\lbrack 4\rbrack_X\lbrack 1111\rbrack_{CFS}\lbrack 211\rbrack_C
\lbrack 31\rbrack_{FS}\lbrack f\rbrack_{F}\lbrack f\rbrack_S$
at the indicated energy.

\begin{center}
\begin{tabular}{|l|c|r|r|} \hline
$\lbrack 31\rbrack_X\lbrack 211\rbrack_{CFS}\lbrack 211\rbrack_C$
& Energy & & \\
$\lbrack f\rbrack_{FS}\lbrack f\rbrack_F\lbrack f\rbrack_S$
& (MeV) & $J^P$ & $\Omega^-$ (emp.) \\
\hline
& & &\\
$\lbrack 4\rbrack_{FS}\lbrack 31\rbrack_F\lbrack 31\rbrack_S$ &
$1865$ & ${1\over2}^+$ &\\
& & &\\
& & ${3\over2}^+$ & \\
& & & \\
& & ${5\over2}^+$ & \\
& & & \\
$\lbrack 31\rbrack_{FS}\lbrack 31\rbrack_F\lbrack 22\rbrack_S$ &
$2145$ & ${1\over2}^+$ & \\
& $+$  & & \\
& & ${3\over2}^+$ & \\
& & & \\
$\lbrack 4\rbrack_{FS}\lbrack 4\rbrack_F\lbrack 4\rbrack_S$ &
$2145$ & ${1\over2}^+$ & $\Omega^-(?)$\\
& & & \\
& & ${3\over2}^+$ & $\Omega^-(?)$ \\
& & & \\
& & ${5\over2}^+$ & $\Omega^-(?)$ \\
& & & \\
& & ${7\over2}^+$ & \\
& & & \\
$\lbrack 31\rbrack_{FS}\lbrack 31\rbrack_F\lbrack 31\rbrack_S$ &
$2201$ & ${1\over2}^+$ & $\Omega^-(2250)$? \\
& $+$ & & \\
& & ${3\over2}^+$ & $\Omega^-(2250)$?\\
& & & \\
& & ${5\over2}^+$ & $\Omega^-(2250)$?\\
& & & \\
\hline
\end{tabular}
\end{center}

\newpage

\centerline{\bf Table 12, continued}

\begin{center}
\begin{tabular}{|l|c|r|r|} \hline
$\lbrack 31\rbrack_X\lbrack 211\rbrack_{CFS}\lbrack 211\rbrack_C$
& Energy & & \\
$\lbrack f\rbrack_{FS}\lbrack f\rbrack_F\lbrack f\rbrack_S$
& (MeV) & $J^P$ & $\Omega^-$ (emp.) \\
\hline
& & &\\
$\lbrack 31\rbrack_{FS}\lbrack 31\rbrack_F\lbrack 4\rbrack_S$ &
$2313$ & ${1\over2}^+$ & \\
& $+$ & & \\
& & ${3\over2}^+$ & \\
& & & \\
& & ${5\over2}^+$ & \\
& & & \\
& & ${7\over2}^+$ & \\
& & & \\
$\lbrack 22\rbrack_{FS}\lbrack 31\rbrack_F\lbrack 31\rbrack_S$ &
$2369$ & ${1\over2}^+$ & \\
& & & \\
& & ${3\over2}^+$ & \\
& & & \\
& & ${5\over2}^+$ & \\
& & & \\
$\lbrack 31\rbrack_{FS}\lbrack 4\rbrack_F\lbrack 31\rbrack_S$ &
$2369$ & ${1\over2}^+$ & \\
& $+$ & & \\
& & ${3\over2}^+$ & \\
& & & \\
& & ${5\over2}^+$ & \\
& & & \\
$\lbrack 4\rbrack_{FS}\lbrack 4\rbrack_F\lbrack 4\rbrack_S$ ($ssss\bar s$)&
$2385$ & ${1\over2}^+$ & $\Omega^-(2380)$(**)? \\
& & & \\
& & ${3\over2}^+$ & $\Omega^-(2380)$(**)? \\
& & & \\
& & ${5\over2}^+$ & $\Omega^-(2380)$(**)? \\
& & & \\
& & ${7\over2}^+$ & $\Omega^-(2380)$(**)? \\
& & & \\
\hline
\end{tabular}
\end{center}

\end{document}